\documentclass[preprintnumbers,10pt,nofootinbib]{revtex4}

\usepackage{amsmath,latexsym,amssymb,amsfonts}
\usepackage[dvips]{graphicx,color}
\usepackage{bm}

\begin{document}


\title{\textbf{Cusp singularities in $f(R)$ gravity: \textit{pros} and \textit{cons}}}

\author{
\textsc{Pisin Chen}$^{a,b,c,d}$\footnote{{\tt pisinchen{}@{}phys.ntu.edu.tw}} and
\textsc{Dong-han Yeom}$^{a}$\footnote{{\tt innocent.yeom{}@{}gmail.com}}
}

\affiliation{
$^{a}$\small{Leung Center for Cosmology and Particle Astrophysics, National Taiwan University, Taipei 10617, Taiwan}\\
$^{b}$\small{Department of Physics, National Taiwan University, Taipei 10617, Taiwan}\\
$^{c}$\small{Graduate Institute of Astrophysics, National Taiwan University, Taipei 10617, Taiwan}\\
$^{d}$\small{Kavli Institute for Particle Astrophysics and Cosmology, SLAC National Accelerator Laboratory, Stanford University, Stanford, California 94305, USA}
}

\begin{abstract}
We investigate cusp singularities in $f(R)$ gravity, especially for Starobinsky and Hu-Sawicki dark energy models. We illustrate that, by using double-null numerical simulations, a cusp singularity can be triggered by gravitational collapses. This singularity can be cured by adding a quadratic term, but this causes a Ricci scalar bump that can be observed by an observer outside the event horizon. Comparing with cosmological parameters, it seems that it would be difficult to see super-Planckian effects by astrophysical experiments. On the other hand, at once there exists a cusp singularity, it can be a mechanism to realize a horizon scale curvature singularity that can be interpreted by a firewall.
\end{abstract}

\maketitle

\newpage

\tableofcontents


\section{Introduction}

It is generally believed that Einstein gravity is not a complete theory and therefore it should be modified. In this regard, the introduction of higher curvature terms has been a natural choice by many researchers, motivated partly by its connection with fundamental concepts such as string theory \cite{BBS,Gasperini:2007zz}. Among various contributions of higher curvature terms, a function of the Ricci scalar is the most simplest extension; such model can be transformed to a scalar-tensor gravity model as well as Einstein gravity via a conformal transformation \cite{ST,Sotiriou:2008rp}. This is commonly referred to as $f(R)$ gravity.

Since $f(R)$ gravity is transformable to a Brans-Dicke type theory \cite{Brans:1961sx}, $f(R)$ gravity is in general easier to control compared to the other general higher curvature corrections. This means that the behavior of the Ricci scalar can be understood from the dynamics of the effective gravitational coupling, i.e., the Brans-Dicke field. Thus, $f(R)$ gravity can provide a natural motivation for the choice of certain shape of a scalar field potential in the Einstein frame that may be useful for an inflationary model \cite{Starobinsky:1980te} or a dark energy model for late time cosmology \cite{Starobinsky:2007hu,Hu:2007nk}.

The dynamics of the Brans-Dicke field will be determined by its potential that depends on the choice of $f(R)$. There is, however, a shortcoming of such an approach. If the shape of $f(R)$ of a model is unwisely chosen, then it may result in a cusp in the potential of the Brans-Dicke field; this may imply the divergence of the Ricci scalar \cite{Appleby:2009uf}. If a region has a divergent Ricci scalar, then it should mean a singularity. However, since this is due to an unwise choice of $f(R)$, such singularity may be fictitious and it may mean that there should be a way to regularize it. We call this as the $f(R)$-induced singularity or the cusp singularity \cite{Hwang:2011kg}.

The follows are known results in the literature:
\begin{itemize}
\item[--] Many dark energy models of $f(R)$ gravity contain a cusp singularity. In order to obtain a stable and non-vanishing (de Sitter) local minimum, one often has to fabricate the shape of the effective potential of the Brans-Dicke field; equivalently, one has to tune a form of $f(R)$ \cite{Starobinsky:2007hu,Hu:2007nk}. For this, the $R^{-n}$ ($n > 0$) corrections have often been introduced, which nevertheless become the origin of the trouble.
\item[--] If the correction term is quadratic, i.e., $\sim R^{2}$, then this cusp singularity can be ameliorated \cite{Appleby:2009uf}. Such a correction term is highly motivated as a viable model for inflation \cite{Starobinsky:1980te,Ade:2015lrj}.
\item[--] If the cusp singularity is inevitable, then it may be realized during a gravitational collapse \cite{Hwang:2011kg,Guo:2013dha}. Furthermore, such a singularity can be located outside the event horizon. Therefore, if it does exist, then it can cause a detectable naked singularity.
\end{itemize}
From these, a question naturally arises as follows. Even though we cure the cusp singularity, there are still possibilities that a gravitational collapse induces a very higher curvature region. Even though it does not diverge, if it reaches around the Planck scale, then it will cause Planck-scale physical effects. Is it really possible and what will be the parameters that determine such a phenomenology? Regarding this, there can be two opinions.
\begin{itemize}
\item[\textit{Cons}:] As long as there is a Planck-scale effect, it can result a distinctive outcome to outside the black hole, while we do not have any evidences of Planck-scale effects during gravitational collapses. Therefore, we should avoid a model that causes a cusp singularity.
\item[\textit{Pros}:] If the existence of a naked singularity is inevitable, then it implies that there appears Planck-scale effects around the horizon scale. This will be useful to understand the unitarity issue of black hole dynamics.
\end{itemize}
The discussion on these two opposite opinions is the motivation of this paper.

In this paper, we study a gravitational collapse of $f(R)$ dark energy models that contain cusp singularities, by using numerical simulations. Especially we use double-null formalism \cite{Hamade:1995ce} to study the formation of charged black holes \cite{Hong:2008mw,doublenull} where this can be applied for cosmology \cite{Hansen:2009kn} as well as scalar-tensor models \cite{Hwang:2010aj,Scheel:1994yr}. We first show that we can cure the cusp singularity problem by adding a $\sim R^{2}$ term; in addition, we show that there can be a higher curvature region.

This paper is organized as follows. In SEC.~\ref{sec:model}, we illustrate our model as well as its realization to double-null formalism. In SEC.~\ref{sec:res}, we discuss the details of numerical results and observe phenomena of cured cusp singularities. Finally, in SEC.~\ref{sec:dis}, we summarize and discuss observational implications and future applications. In this paper, we use the natural units: $c = G = \hbar = 1$.

\section{\label{sec:model}Double-null formalism for $f(R)$ gravity}

In this section, we discuss the way to realize double-null formalism for $f(R)$ gravity models. We follow previous notations of \cite{Hong:2008mw}. To realize $f(R)$ gravity to double-null formalism, we especially used a technique of \cite{Hwang:2011kg}. Mathematical details are in Appendix A and its convergence and consistency checks are discussed in Appendix B.

\subsection{$f(R)$ gravity}

The action of the $f(R)$ gravity model is presented by
\begin{eqnarray}
S = \int d^{4}x \sqrt{-g} \left[ \frac{1}{16\pi} f(R) + \mathcal{L}_{\mathrm{matter}} \right],
\end{eqnarray}
where $R$ is the Ricci scalar and $f(R)$ is a given function of $R$. By introducing an auxiliary field, this model can be transformed to the Brans-Dicke type model \cite{ST}:
\begin{eqnarray}
S = \int d^{4}x \sqrt{-g} \left[\frac{1}{16\pi} \left( \Phi R - V(\Phi) \right) + \mathcal{L}_{\mathrm{matter}}\right],
\end{eqnarray}
where $V(\Phi) = - f(R) + R f'(R)$ and $\Phi = f'(R)$. This is the $\omega = 0$ limit of the Brans-Dicke theory with a potential $V(\Phi)$.

In order to induce a gravitational collapse, one convenient way is to introduce a gauge field, $A_{\mu}$, that couples to a matter field, for which we consider a complex scalar field $\phi$. To implement this, we use the following model \cite{Hong:2008mw,Hwang:2010aj}:
\begin{eqnarray}\label{eq:BDscalar}
S_{\mathrm{BD}} = \int d^{4}x \sqrt{-g} \left[ \frac{1}{16\pi} \left( \Phi R - \frac{\omega}{\Phi}\Phi_{;\mu}\Phi_{;\nu}g^{\mu\nu} - V(\Phi) \right) + \mathcal{L}^{\mathrm{EM}} \right],
\end{eqnarray}
where
\begin{eqnarray}
\mathcal{L}^{\mathrm{EM}} = - \frac{1}{2}\left(\phi_{;\mu}+ieA_{\mu}\phi \right)g^{\mu\nu}\left(\overline{\phi}_{;\nu}-ieA_{\nu}\overline{\phi}\right)-\frac{1}{16\pi}F_{\mu\nu}F^{\mu\nu},
\end{eqnarray}
$e$ is a gauge coupling and $F_{\mu\nu}=A_{\nu;\mu}-A_{\mu;\nu}$. We choose $\omega=0$ throughout this paper.

The Einstein equation is as follows:
\begin{eqnarray}\label{eq:Einstein}
G_{\mu\nu} = 8 \pi T^{\mathrm{BD}}_{\mu\nu} + 8 \pi \frac{T^{\mathrm{C}}_{\mu\nu}}{\Phi} \equiv 8 \pi T_{\mu\nu},
\end{eqnarray}
where the Brans-Dicke part of the energy-momentum tensor is
\begin{eqnarray}\label{eq:T_BD}
T^{\mathrm{BD}}_{\mu\nu} = \frac{1}{8\pi \Phi} \left(-g_{\mu\nu}\Phi_{;\rho \sigma}g^{\rho\sigma}+\Phi_{;\mu\nu}\right)
+ \frac{\omega}{8\pi \Phi^{2}} \left(\Phi_{;\mu}\Phi_{;\nu}-\frac{1}{2}g_{\mu\nu}\Phi_{;\rho}\Phi_{;\sigma}g^{\rho\sigma}\right) - g_{\mu\nu}\frac{V(\Phi)}{16 \pi}
\end{eqnarray}
and the matter part of the energy-momentum tensor is
\begin{eqnarray}\label{eq:T_C}
T^{\mathrm{C}}_{\mu\nu} &=& \frac{1}{2}\left(\phi_{;\mu}\overline{\phi}_{;\nu}+\overline{\phi}_{;\mu}\phi_{;\nu}\right)+\frac{1}{2}\left(-\phi_{;\mu}ieA_{\nu}\overline{\phi}+\overline{\phi}_{;\nu}ieA_{\mu}\phi+\overline{\phi}_{;\mu}ieA_{\nu}\phi-\phi_{;\nu}ieA_{\mu}\overline{\phi}\right)
\nonumber \\
&& {}+\frac{1}{4\pi}F_{\mu \rho}{F_{\nu}}^{\rho}+e^{2}A_{\mu}A_{\nu}\phi\overline{\phi}+\mathcal{L}^{\mathrm{EM}}g_{\mu \nu}.
\end{eqnarray}
Field equations for $\Phi$, $\phi$, and $A_{\mu}$ are as follows:
\begin{eqnarray}
\label{eq:Phi1}\Phi_{;\mu\nu}g^{\mu\nu}-\frac{8\pi}{3+2\omega}T^{\mathrm{C}}-\frac{1}{3+2\omega}\left( \Phi \frac{dV}{d\Phi} - 2 V \right) &=&0, \\
\label{eq:phi}\phi_{;\mu\nu}g^{\mu\nu}+ieA^{\mu}\left(2\phi_{;\mu}+ieA_{\mu}\phi\right)+ieA_{\mu;\nu}g^{\mu\nu}\phi &=& 0,
\\
\label{eq:a}\frac{1}{2\pi} {F^{\nu}}_{\mu;\nu} -ie\phi\left(\overline{\phi}_{;\mu}-ieA_{\mu}\overline{\phi}\right)+ie\overline{\phi}\left(\phi_{;\mu}+ieA_{\mu}\phi\right) &=& 0,
\end{eqnarray}
where $T^{\mathrm{C}} = {T^{\mathrm{C}}}^{\mu}_{\;\mu}$. For convenience, we define the effective potential $U(\Phi)$ by
\begin{eqnarray}
U(\Phi) = \int^{\Phi} \left( \bar{\Phi} V'(\bar{\Phi}) - 2V(\bar{\Phi}) \right) d \bar{\Phi},
\end{eqnarray}
where this determines the dynamics of the field.

\subsection{\label{sec:ini}Initial conditions and free parameters}

We use the double-null coordinate
\begin{eqnarray}\label{eq:doublenull}
ds^{2} = -\alpha^{2}(u,v) du dv + r^{2}(u,v) d\Omega^{2}
\end{eqnarray}
for numerical simulations. Here, $u$ is the retarded time, $v$ is the advanced time, and $d\Omega^{2}$ is the two-sphere. For numerical integration, it is convenient to change all equations by a set of first order differential equations. For this purpose, we introduce the following notations \cite{Hwang:2011kg,Hong:2008mw,Hansen:2009kn,Hwang:2010aj}: the metric function $\alpha$, the radial function $r$, the Brans-Dicke field $\Phi$, and a scalar field $s \equiv \sqrt{4\pi} \phi$, and define
\begin{eqnarray}\label{eq:conventions}
h \equiv \frac{\alpha_{,u}}{\alpha},\quad d \equiv \frac{\alpha_{,v}}{\alpha},\quad f \equiv r_{,u},\quad g \equiv r_{,v},\quad W \equiv \Phi_{,u},\quad Z \equiv \Phi_{,v}, \quad w \equiv s_{,u},\quad z \equiv s_{,v}.
\end{eqnarray}
In addition, in order to avoid the complicated relation between $f(R)$ and $V(\Phi)$, we use the evolution of Ricci scalars (see Appendix A). Then we need initial conditions for all functions ($\alpha, h, d, r, f, g, \Phi, W, Z, s, w, z, a, q, R$) at the initial $u=u_{\mathrm{i}}$ and $v=v_{\mathrm{i}}$ surfaces, where we set $u_{\mathrm{i}}=v_{\mathrm{i}}=0$.

We have a gauge freedom to choose the initial $r$ function, i.e., we have a freedom to choose distances between null lines. We choose $r(0,0)=r_{0}$, $f(u,0)=r_{u0}$, and $g(0,v)=r_{v0}$, where $r_{u0}<0$ and $r_{v0}>0$ in order to implement that $r$ increases for an out-going direction and decreases for an in-going direction. We assume that $\Phi$ is asymptotically a constant $\Phi_{0}$ at a local minimum of the effective potential, so that $\Phi_{0} V'(\Phi_{0}) - 2 V(\Phi_{0}) = 2 f(R_{0})-R_{0}f'(R_{0})=0$. Then, $\Phi(u,0)=\Phi(0,v)=\Phi_{0}$, $W(u,0)=Z(0,v)=0$, and $R(u,0)=R(0,v)=R_{0}$. In addition, we need the following information.
\begin{itemize}
\item[--] \textit{In-going null direction}: We choose $s(u,0)=0$ and $w(u,0)=h(u,0)=a(u,0)=q(a,0)=0$. Since the asymptotic Misner-Sharp mass function
\begin{eqnarray}
m(u,v) \equiv \frac{r}{2}\left( 1+4\frac{r_{,u}r_{,v}}{\alpha^{2}} + \frac{q^{2}}{r^{2}} - \frac{V}{6\Phi}r^{2} \right)
\end{eqnarray}
should vanish at $u=v=0$, it is convenient to choose $r_{u0}=-1/2$ and $r_{v0}=1/2$, and
\begin{eqnarray}
\alpha(0,0)=\left(1-\frac{V(\Phi_{0})}{6\Phi_{0}} r_{0}^{2}\right)^{-1/2}.
\end{eqnarray}
We need more information to determine $d, g, Z$, and $z$ at the $v=0$ surface. We obtain $d$ from Eq.~(\ref{eq:E3}), $g$ from Eq.~(\ref{eq:E4}), $Z$ from Eq.~(\ref{eq:Phi}), and $z$ from Eq.~(\ref{eq:fieldeqns3}).
\item[--] \textit{Out-going null dierction}: In order to derive a gravitational collapse, we need to choose a shape of an in-going scalar field $s(0,v)$. In principle, we can choose this scalar field profile arbitrarily, but for convenience, we use
\begin{eqnarray} \label{s_initial}
s(u_{\mathrm{i}},v)= A \sin^{2} \left( \pi \frac{v-v_{\mathrm{i}}}{v_{\mathrm{f}}-v_{\mathrm{i}}} \right) \left[ \cos \left( 2 \pi \frac{v-v_{\mathrm{i}}}{v_{\mathrm{f}}-v_{\mathrm{i}}} \right) + i \cos\left( 2 \pi \frac{v-v_{\mathrm{i}}}{v_{\mathrm{f}}-v_{\mathrm{i}}} + \delta \right) \right]
\end{eqnarray}
for $0\leq v \leq v_{\mathrm{f}}$ and $s(0,v)=0$ otherwise, where $v_{\mathrm{f}}$ is the width of the pulse, $A$ is the amplitude, and $\delta$ is a free parameter to tune the phase of the complex scalar field \cite{Hwang:2011kg,Hong:2008mw}. Taking the derivative with respect to $v$, we obtain $z(u_{\mathrm{i}},v)$. From Eq.~(\ref{eq:E2}), we find $d = r|z|^{2}/2g\Phi$ on the $u=0$ surface. By integrating $d$ along $v$, we also get $\alpha(0,v)$. However, we need more information for $h, f, W, w, a,$ and $q$ at the $u=0$ surface. As shown in Appendix A, we obtain $h$ from Eq.~(\ref{eq:E3}), $f$ from Eq.~(\ref{eq:E4}), $W$ from Eq.~(\ref{eq:Phi}), $w$ from Eq.~(\ref{eq:fieldeqns3}), $a$ from Eq.~(\ref{eq:fieldeqns1}), and $q$ from Eq.~(\ref{eq:fieldeqns2}).
\end{itemize}
Finally, we choose $r_{0}=10$, $\delta = 0.5 \pi$, $\omega=0$, $v_{\mathrm{f}} = 20$, and leaving $A$ and $e$ free.

\begin{figure}
\begin{center}
\includegraphics[scale=0.35]{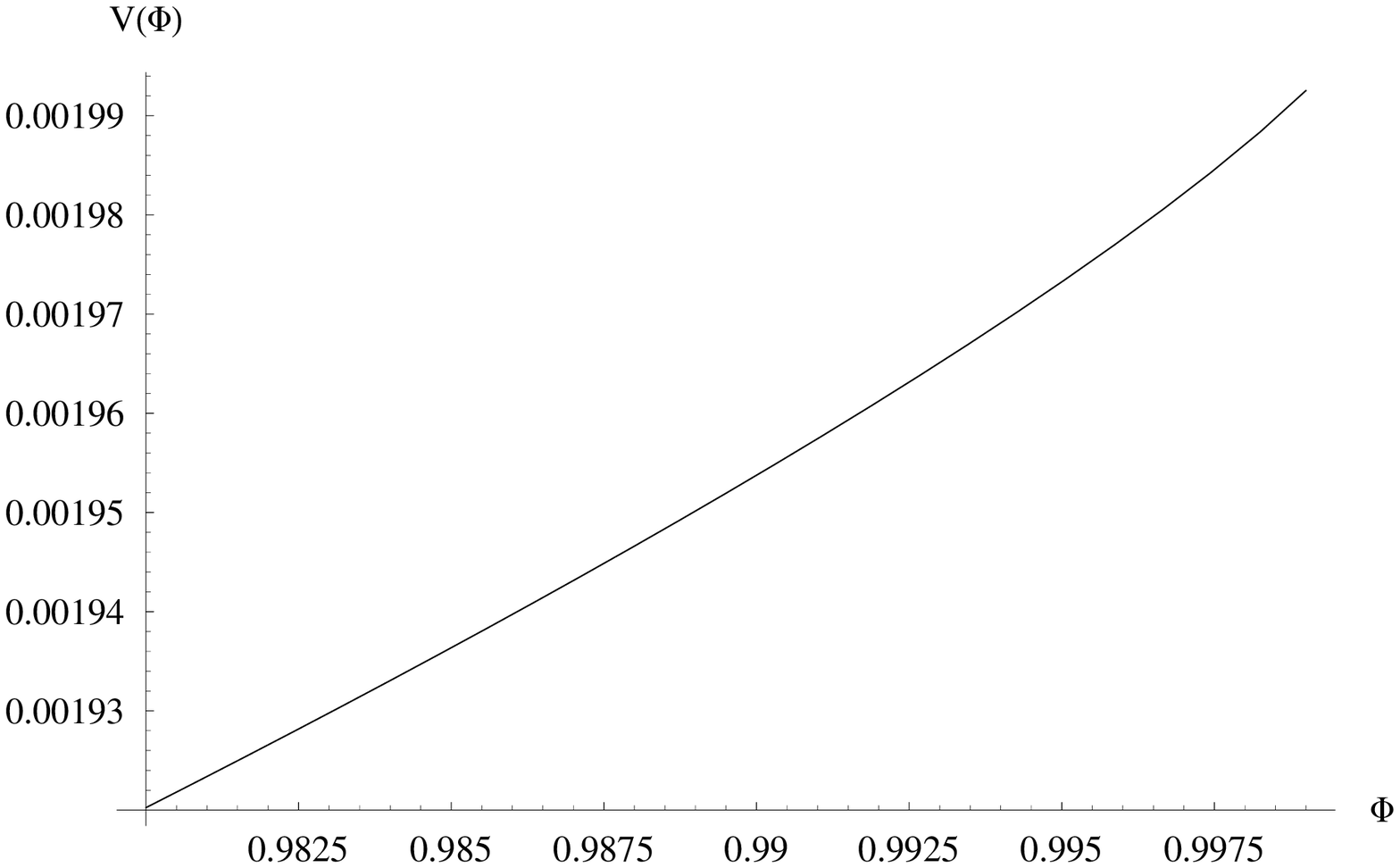}
\includegraphics[scale=0.35]{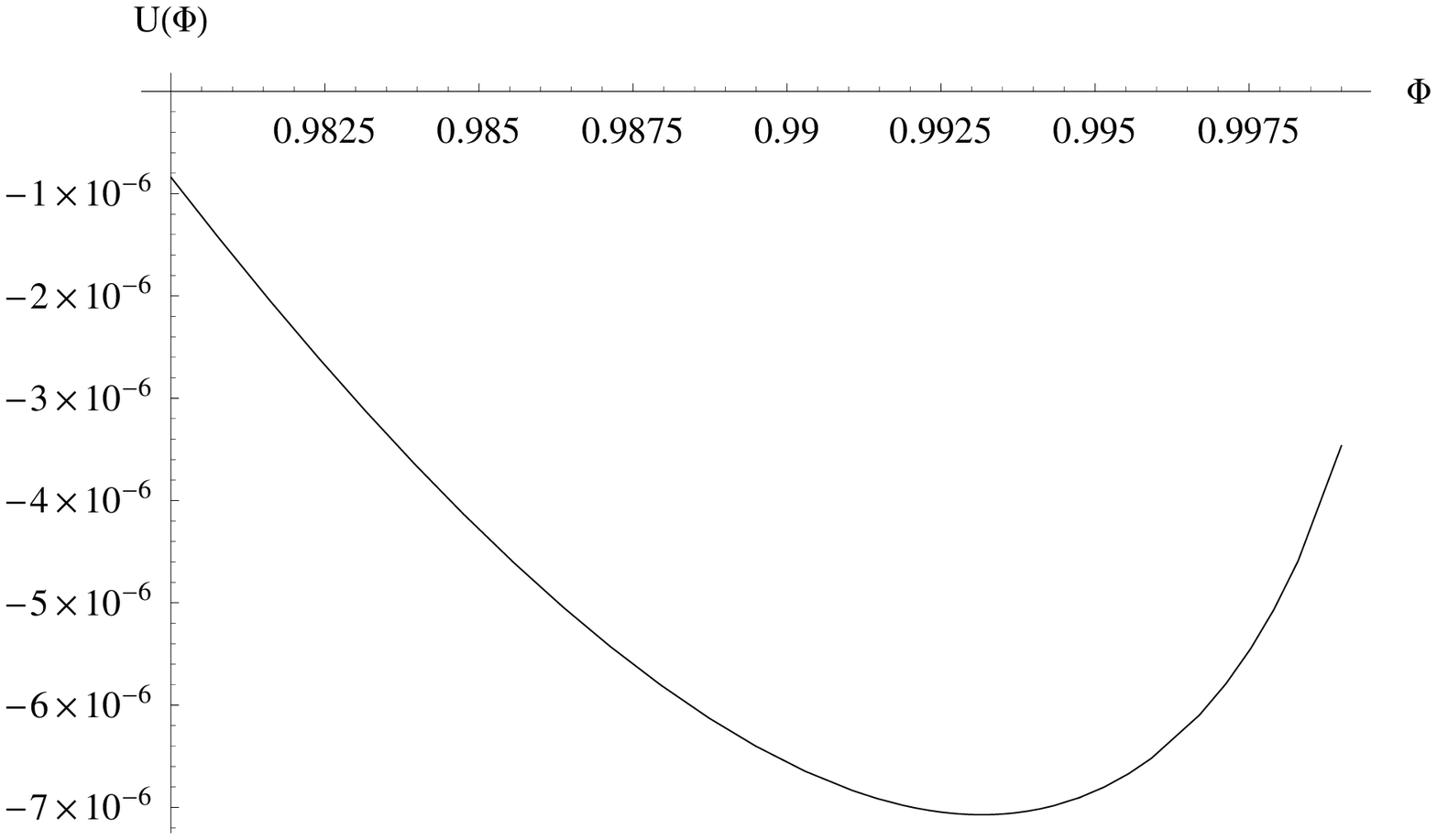}
\caption{\label{fig:potential_S}$V(\Phi)$ and $U(\Phi)$ for $f_{\mathrm{S}}(R)$.}
\includegraphics[scale=0.35]{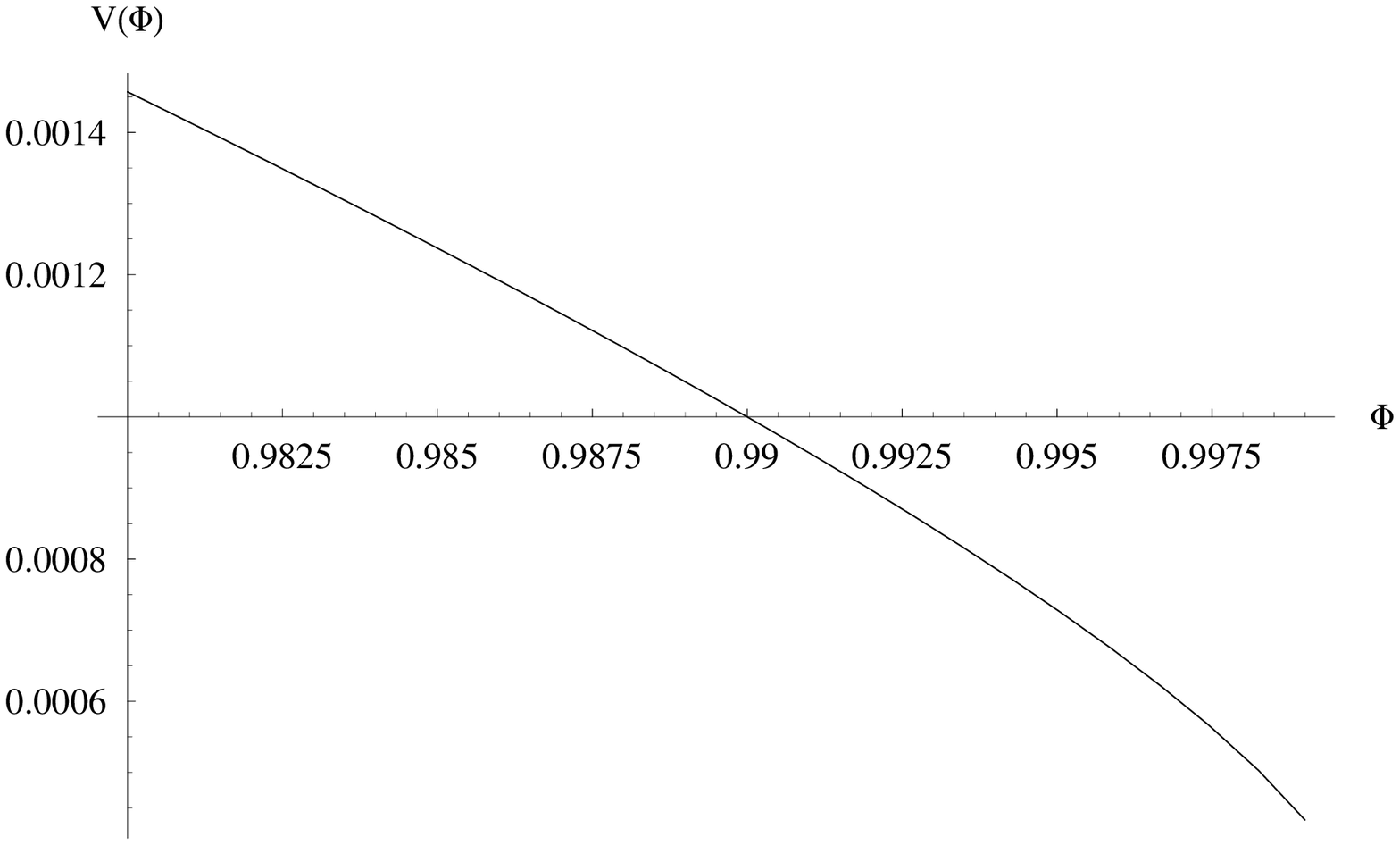}
\includegraphics[scale=0.35]{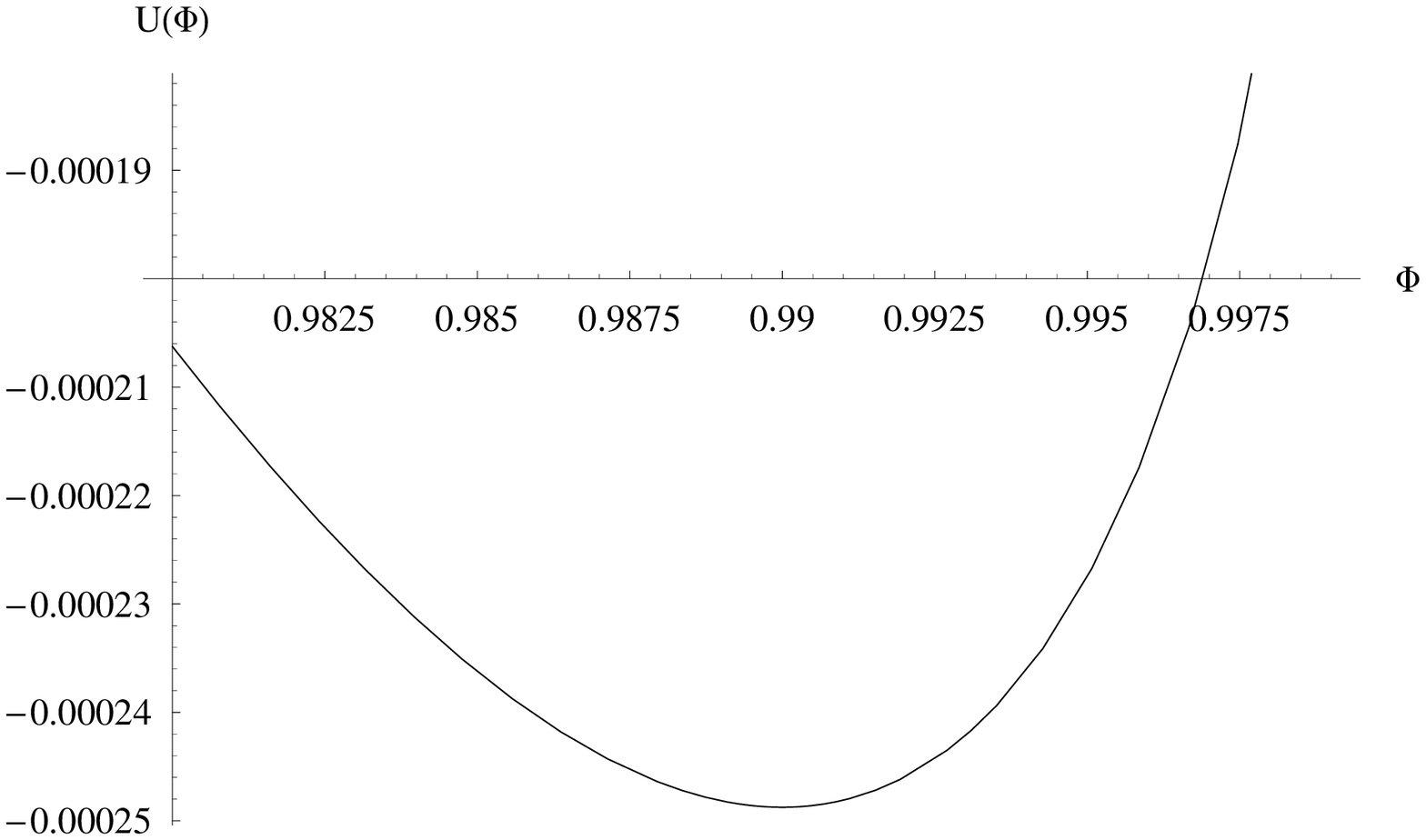}
\caption{\label{fig:potential_HS}$V(\Phi)$ and $U(\Phi)$ for $f_{\mathrm{HS}}(R)$.}
\end{center}
\end{figure}

\subsection{\label{sec:Grav}Choice of $f(R)$: cusp singularities and their remedies}

In addition to the initial conditions, we need to specify the detailed models of $f(R)$. In this paper, we are especially interested in four models: the Starobinsky (dark energy) model \cite{Starobinsky:2007hu}, the Hu-Sawicki model \cite{Hu:2007nk}, and their cured forms by adding $\sim R^{2}$ corrections.

\paragraph{Starobinsky model}

The Starobinsky (dark energy) model is as follows \cite{Starobinsky:2007hu}:
\begin{eqnarray}
f_{\mathrm{S}}(R) = R + \lambda R_{0} \left[ \left(1+\frac{R^{2}}{R_{0}^{2}} \right)^{-n}-1 \right],
\end{eqnarray}
where $\lambda$, $R_{0}$, and $n$ are free parameters. We choose $\lambda=2$, $R_{0}=0.001$, and $n=2$ (FIG.~\ref{fig:potential_S}). We can see that there is a stable equilibrium around $\Phi_{0}\simeq0.9932$ and $V_{0}\simeq0.001966$.

In the large $R$ limit, by using the Taylor expansion, we observe
\begin{eqnarray}
f_{\mathrm{S}}(R) &\simeq& R - \lambda R_{0} + \lambda R_{0}^{2n+1} \frac{1}{R^{2n}} + ...,\\
f'_{\mathrm{S}}(R) &\simeq& 1 - 2n \lambda R_{0}^{2n+1} \frac{1}{R^{2n+1}} + ...,\\
f''_{\mathrm{S}}(R) &\simeq& 2n(2n+1) \lambda R_{0}^{2n+1} \frac{1}{R^{2n+2}} + ...,\\
\lim_{\Phi \rightarrow 1} U(\Phi) &=& \lim_{\Phi \rightarrow 1} \int^{\Phi} \left(\bar{\Phi} \frac{V(\bar{\Phi})}{d \bar{\Phi}} - 2V(\bar{\Phi}) \right) d\bar{\Phi}\\
&=& \lim_{R \rightarrow \infty} \int^{R} \left(2f_{\mathrm{S}}(\bar{R}) - \bar{R} f'_{\mathrm{S}}(\bar{R}) \right) f''_{\mathrm{S}}(\bar{R}) d\bar{R}\\
&\propto& \int^{\infty} \frac{d\bar{R}}{\bar{R}^{2n+1}}.
\end{eqnarray}
If $n>0$, then the effective potential $U(\Phi)$ of the $\Phi=1$ limit is finite, while the effective force term $2f - Rf' = \Phi V' -2 V$ (hence, the limit where the Ricci scalar diverges) is infinite. This shows the possibility to reach the cusp singularity.

\paragraph{Hu-Sawicki model}

We second consider the Hu-Sawicki model \cite{Hu:2007nk}:
\begin{eqnarray}
f_{\mathrm{HS}}(R) = R - \lambda \left[\frac{C_{1}R^{n}}{C_{2} R^{n} + \lambda^{n}} \right].
\end{eqnarray}
We choose $\lambda=1$, $C_{1}=0.01$, $C_{2}=40$, and $n=1$ (FIG.~\ref{fig:potential_HS}). We can see that there is a stable equilibrium around $\Phi_{0}\simeq0.99$ and $V_{0}\simeq0.001$.

In the large $R$ limit, by defining $\bar{\lambda} = \lambda/C_{2}^{1/n}$, we observe
\begin{eqnarray}
f_{\mathrm{HS}}(R) &\simeq& R - \frac{\lambda C_{1}}{C_{2}} + \frac{\lambda C_{1}}{C_{2}} \frac{\bar{\lambda}^{n}}{R^{n}} + ...,\\
f'_{\mathrm{HS}}(R) &\simeq& 1 - n \frac{\lambda C_{1}}{C_{2}} \frac{\bar{\lambda}^{n}}{R^{n+1}} + ...,\\
f''_{\mathrm{HS}}(R) &\simeq& n (n+1) \frac{\lambda C_{1}}{C_{2}} \frac{\bar{\lambda}^{n}}{R^{n+2}} + ...,\\
\lim_{\Phi \rightarrow 1} U(\Phi) &\propto& \int^{\infty} \frac{d\bar{R}}{\bar{R}^{n+1}}.
\end{eqnarray}
Therefore, as was in the Starobinsky model, if $n>0$, there appears a cusp singularity.

\paragraph{Cure of the cusp singularity by inserting $(1/2)cR^{2}$}

In order to avoid the $f(R)$-induced singularity and to study the general feature of the $f(R)$ gravity, we add a correction term that can cure the cusp singularity \cite{Appleby:2009uf}. Here, we study the following `cured' models\footnote{Here, we abuse the parameter $c$ as a constant.}:
\begin{eqnarray}
f_{\mathrm{cure,S}}(R) &=& f_{\mathrm{S}}(R) + \frac{1}{2}c R^{2},\\
f_{\mathrm{cure,HS}}(R) &=& f_{\mathrm{HS}}(R) + \frac{1}{2}c R^{2},
\end{eqnarray}
where $c$ is an arbitrary constant. If $R$ is large enough, then
\begin{eqnarray}
f(R) \simeq R + \frac{1}{2}c R^{2}.
\end{eqnarray}
Since $f'(\psi) \simeq 1 + c \psi = \Phi$, we obtain
\begin{eqnarray}
V(\Phi) \simeq U(\Phi) \simeq \frac{1}{2c} \left( \Phi - 1 \right)^{2},
\end{eqnarray}
and hence there is no cusp singularity now. For stability, we need to choose $c>0$. Throughout this paper, we leave $c$ as a constant.

\section{\label{sec:res}Results and interpretations}

In this section, we first report numerical results on cusp singularities and their regularizations. Second, we discuss the pros and cons of their existence and the possibility of their observations.

\subsection{Numerical results}

\begin{figure}
\begin{center}
\includegraphics[scale=0.3]{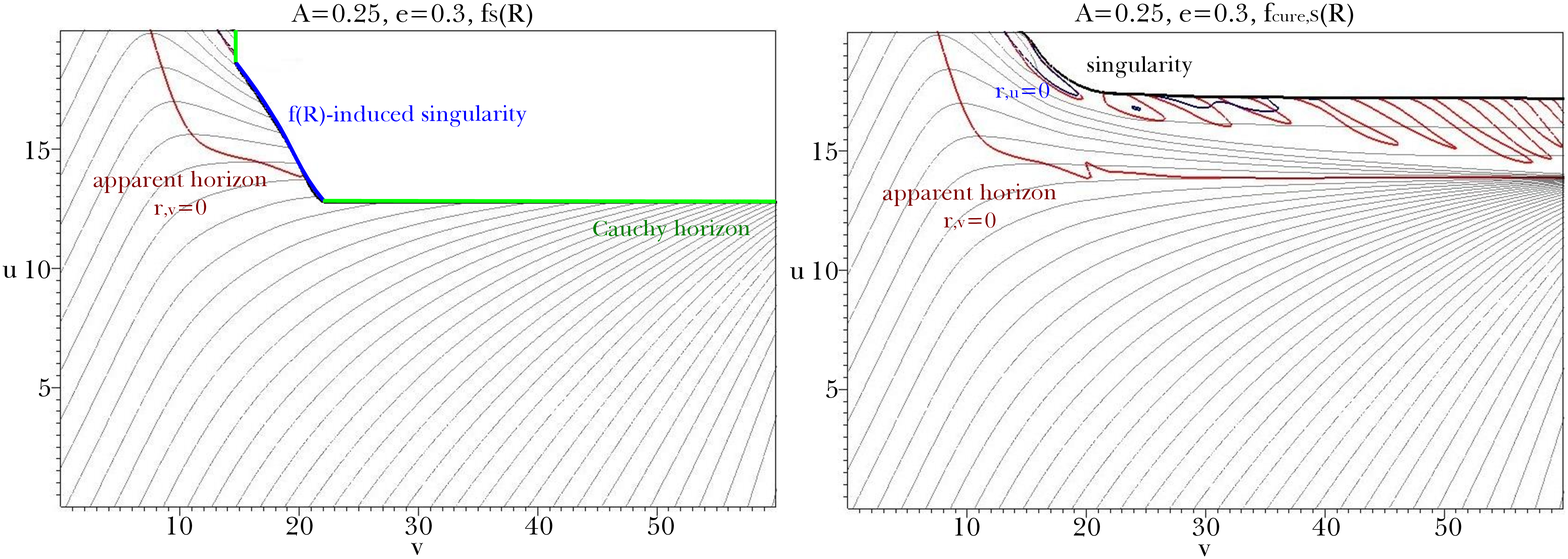}
\includegraphics[scale=0.3]{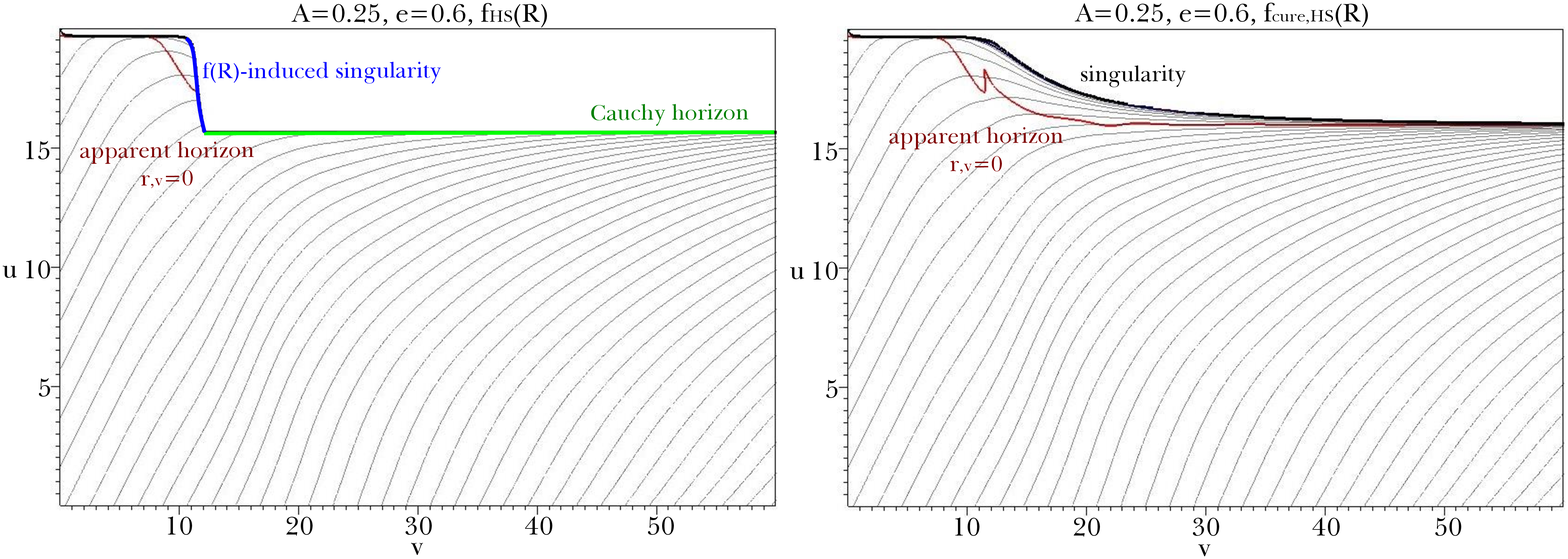}
\caption{\label{fig:causal}Upper: causal structures for $f_{\mathrm{S}}(R)$ and $f_{\mathrm{cure,S}}(R)$ with $A=0.25$, $e = 0.3$, and $c=0.001$. Lower: causal structures for $f_{\mathrm{HS}}(R)$ and $f_{\mathrm{cure,HS}}(R)$ with $A=0.25$, $e = 0.6$, and $c=0.001$.}
\end{center}
\end{figure}
\begin{figure}
\begin{center}
\includegraphics[scale=0.3]{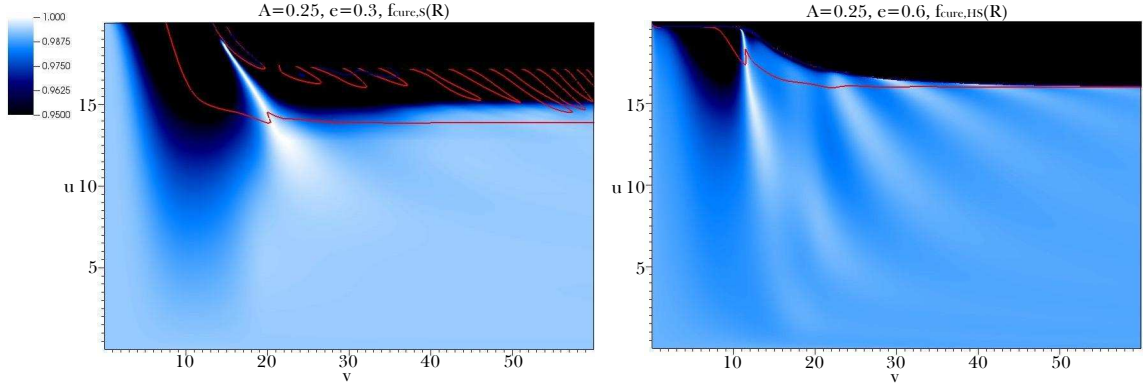}
\caption{\label{fig:Sb}$\Phi$ for $f_{\mathrm{cure,S}}(R)$ with ($A=0.25$, $e = 0.3$, $c=0.001$) and $f_{\mathrm{cure,HS}}(R)$ with ($A=0.25$, $e = 0.6$, $c=0.001$).}
\end{center}
\end{figure}

\paragraph{Numerical results} FIG.~\ref{fig:causal} shows the cases of naked singularities in $f(R)$ dark energy models\footnote{In this paper, we used $\delta = 0.5 \pi$, which this is the case when the charge-mass ratio of a black hole ($Q/M$) is maximum \cite{Hong:2008mw}. There is a tendency that as the charge-mass ratio decreases, the effect of the cusp singularity decreases and the cusp singularity may appear inside the event horizon. For details, see \cite{Hwang:2011kg}.} (upper left is for $f_{\mathrm{S}}(R)$ and lower left is for $f_{\mathrm{HS}}(R)$). However, after one includes the $R^{2}$ regularization term, one can extend beyond the cusp singularity (upper right is for $f_{\mathrm{cure,S}}(R)$ and lower right is for $f_{\mathrm{cure,HS}}(R)$). Before the singularity, the causal structures of $f_{\mathrm{cure,S}}(R)$ and $f_{\mathrm{cure,HS}}(R)$ are almost the same as the causal structures of $f_{\mathrm{S}}(R)$ and $f_{\mathrm{HS}}(R)$. However, around the would-be cusp singularity, the causal structures show some difference between original and regularized models. One typical behavior of the regularized causal structure is that, soon after the would-be cusp singularity, the apparent horizon begins to oscillate. This is due to the dynamics of the Brans-Dicke field: FIG.~\ref{fig:Sb} explicitly shows $\Phi$ for the cured cases. Without the $R^{2}$ regularization term, we would have $\Phi = 1$ at $R = \infty$. On the other hand, after we add the $R^{2}$ term, $\Phi = 1$ is no more a curvature singularity. One can clearly see that the Brans-Dicke field becomes larger and larger around the would-be cusp singularity. Due to the large perturbations of the Brans-Dicke field, as it goes to the local minimum, there are oscillatory effects of the field (as well as energy-momentum tensors) and this makes the apparent horizon oscillatory.

\begin{figure}
\begin{center}
\includegraphics[scale=0.25]{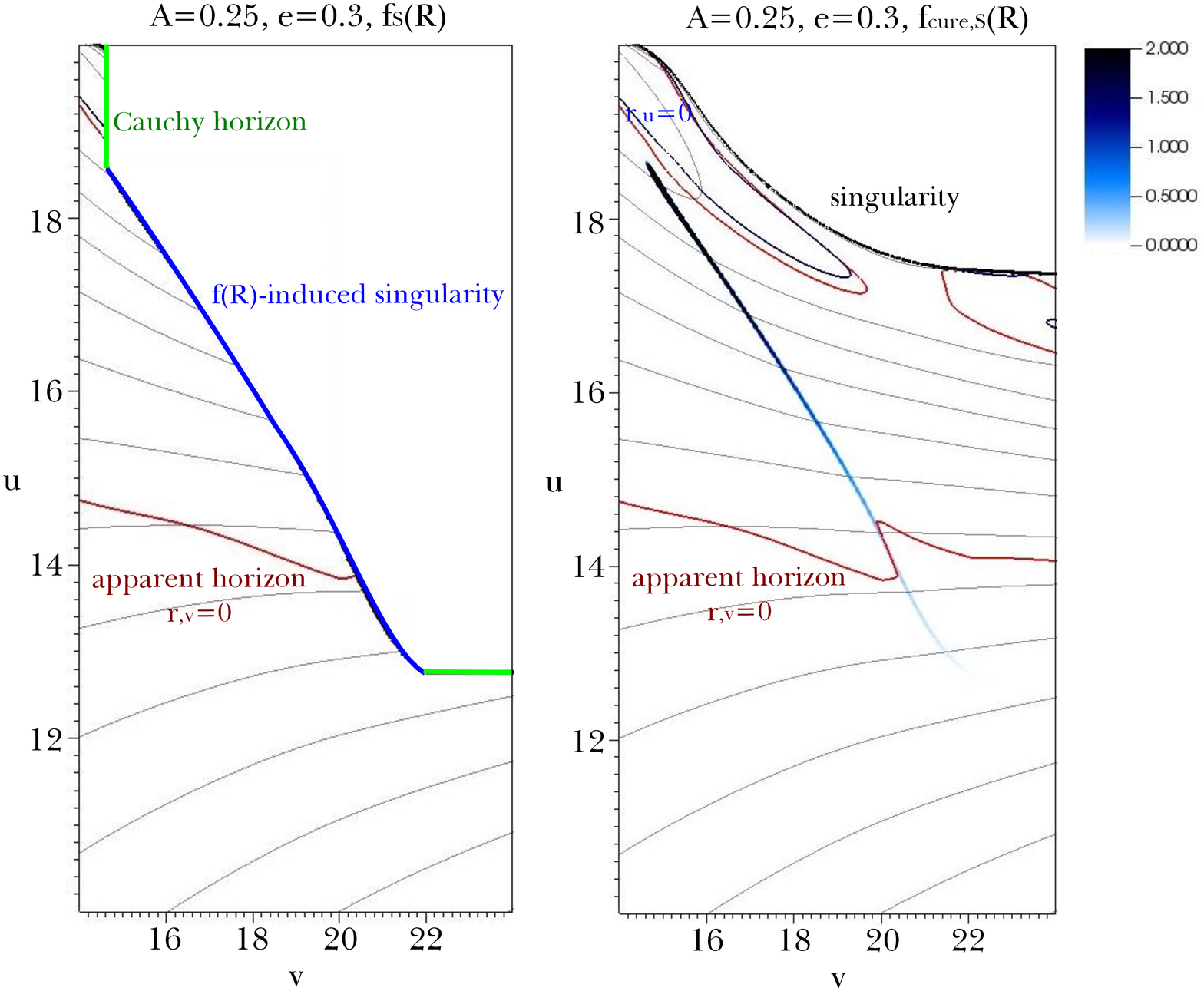}
\includegraphics[scale=0.242]{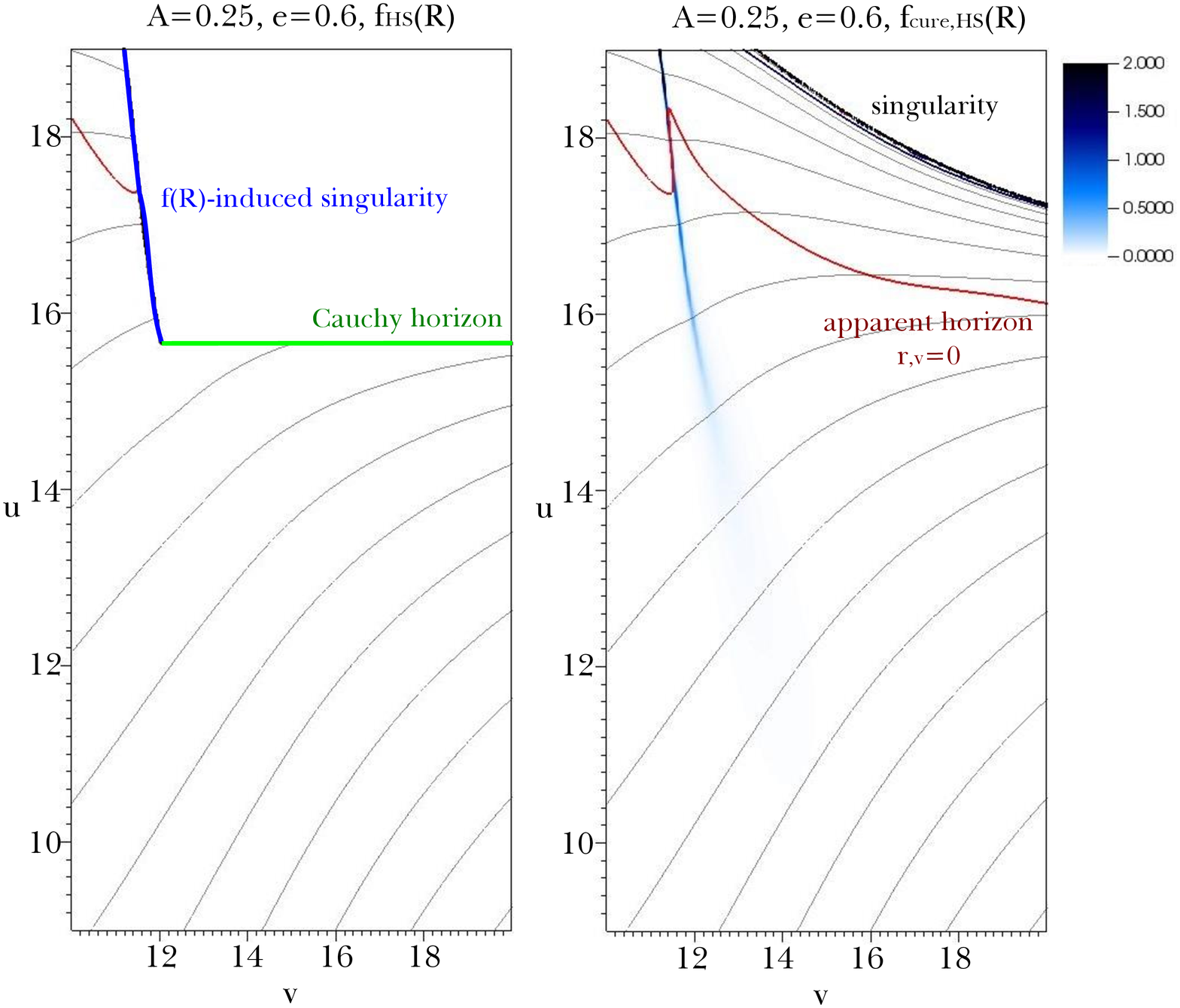}
\caption{\label{fig:Ricci}Left: the causal structure for $f_{\mathrm{S}}(R)$ with ($A=0.25$, $e = 0.3$) and the Ricci scalar $R$ for $f_{\mathrm{cure,S}}(R)$ with ($A=0.25$, $e = 0.3$, $c=0.001$). Right: the causal structure for $f_{\mathrm{HS}}(R)$ with ($A=0.25$, $e = 0.6$) and the Ricci scalar $R$ for $f_{\mathrm{cure,HS}}(R)$ with ($A=0.25$, $e = 0.6$, $c=0.001$). In these figures, in the white-colored region, the Ricci scalar is close to zero, while it becomes relatively larger values around the blue-colored region.}
\end{center}
\end{figure}

\paragraph{Ricci scalar bump} Another salient feature is the behavior of the Ricci scalar. If there is no $R^{2}$ correction term, then the Ricci scalar reaches infinity. However, due to the $R^{2}$ correction term, the Ricci scalar becomes bounded. The point of FIG.~\ref{fig:Ricci} is that, even though the Ricci scalar is bounded, it can increase to a sufficiently large value. The blue colored strips in FIG.~\ref{fig:Ricci} show such behaviors. The Ricci scalar can become larger than the Planck scale ($\sim 1$ in our setup), which can in principle be detected by an observer outside the black hole.

\begin{figure}
\begin{center}
\includegraphics[scale=0.25]{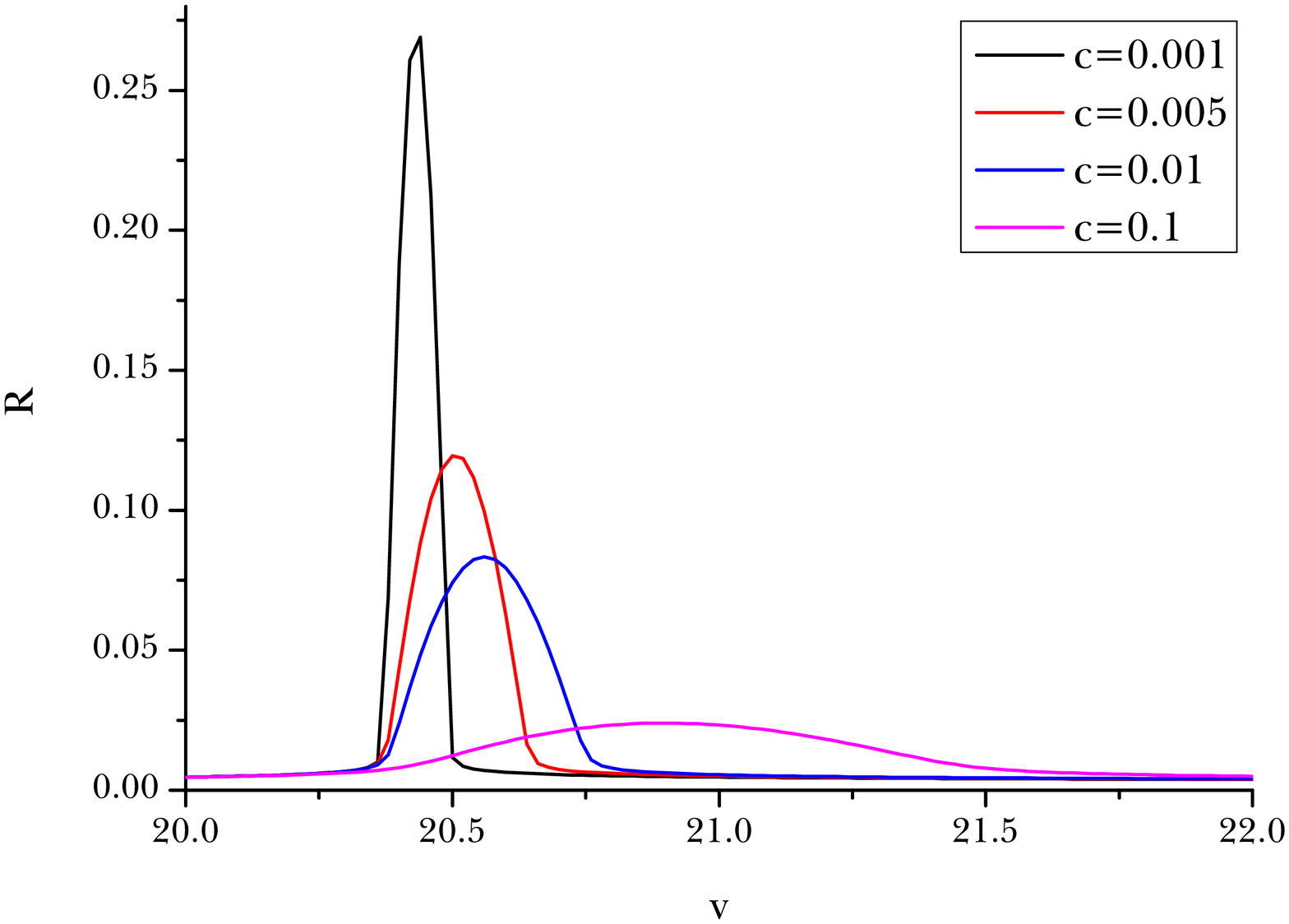}
\includegraphics[scale=0.25]{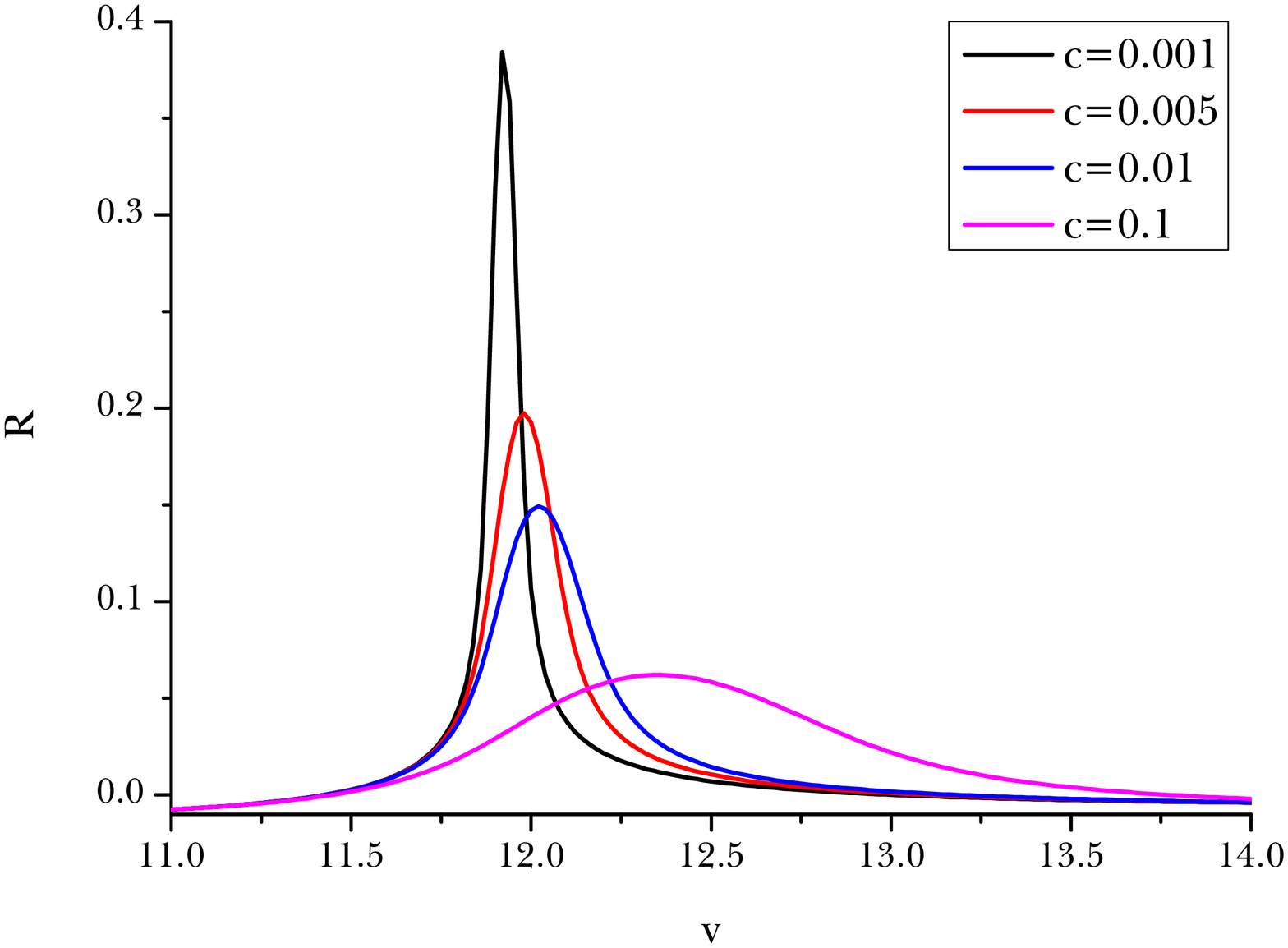}
\caption{\label{fig:eventhorizonRicci}Ricci scalar on the event horizon (as functions of $v$) for $f_{\mathrm{cure,S}}(R)$ (left) and $f_{\mathrm{cure,HS}}(R)$ (right), by varying $c = 0.001, 0.005, 0.01, 0.1$.}
\end{center}
\end{figure}

\begin{figure}
\begin{center}
\includegraphics[scale=0.25]{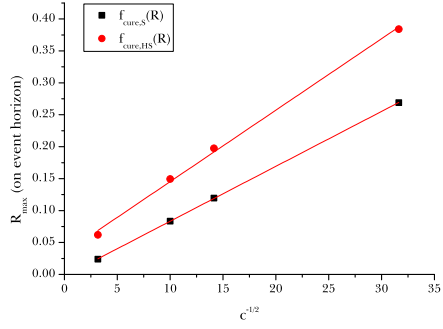}
\caption{\label{fig:fitting}The linear dependence between $1/\sqrt{c}$ and $R_{\mathrm{max}}$ on the event horizon.}
\end{center}
\end{figure}

We can further investigate the size of the Ricci scalar bump by varying the parameter $c$. Although there is no canonical way to define the size of the bump, we may appreciate the maximum value of the Ricci scalar through the event horizon\footnote{In principle, the event horizon is not possible to determine unless the future infinity is known. However, the apparent horizon approaches to a null direction and this will be the event horizon if there is no incoming/outgoing matter around the horizon. In this regard, we determine the location of the event horizon from the asymptotic null direction of the apparent horizon.}. Even though we vary $c$, its back-reaction to the asymptotic mass remains unchanged, and hence the location of the event horizon is not appreciably shifted (though the apparent horizon is sensitively changed by varying $c$). FIG.~\ref{fig:eventhorizonRicci} summarizes our results: as $c$ increases, the maximum value of the Ricci scalar decreases, where the dependence is proportional to $\sim 1/\sqrt{c}$ (FIG.~\ref{fig:fitting}).

\begin{figure}
\begin{center}
\includegraphics[scale=0.6]{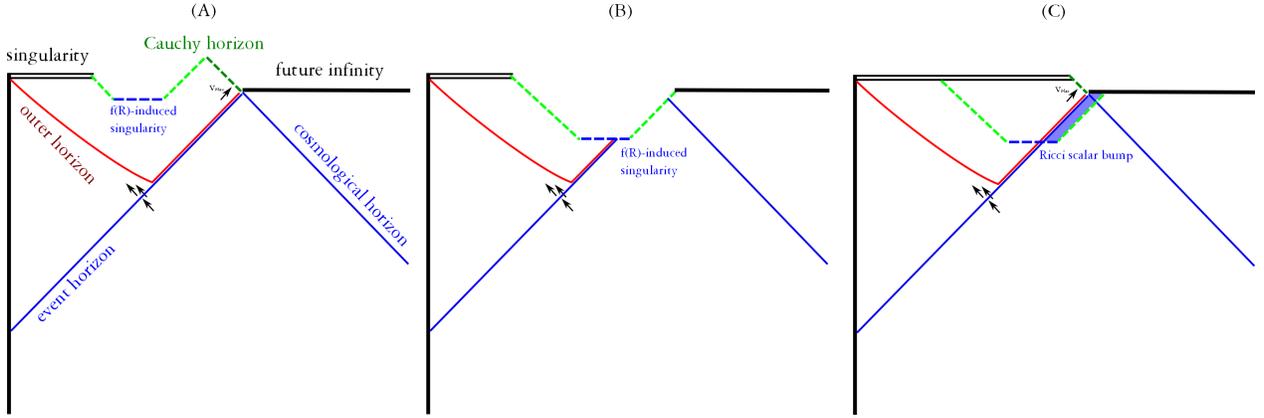}
\caption{\label{fig:diagram}(A) $f(R)$-induced cusp singularity is inside the event horizon \cite{Hwang:2011kg}. (B) As the mass-charge ratio increases, the cusp singularity can be outside the event horizon. (C) If a $\sim R^{2}$ term is added, then the cusp singularity can be removed but a Ricci scalar bump would appear. This can be located outside the horizon and hence can be observed by the outside observer (blue-colored region).}
\end{center}
\end{figure}

\paragraph{Causal structures} FIG.~\ref{fig:diagram} is the summary of the causal structure of gravitational collapses in $f(R)$ gravity. (A) shows the case when the cusp singularity is inside the event horizon. (B) is the case when the cusp singularity is outside the event horizon. If one inserts the $R^{2}$ regularization term, then one can extend the causal structure as denoted in (C). However, due to the Ricci scalar bump, the high curvature (perhaps, super-Planckian) effects can in principle be observed outside the event horizon (in the blue shaded region).

\subsection{Cons: can we see super-Planckian effects for realistic cosmological models?}

In this subsection, let us try a crude estimation of the Ricci scalar bump. If $U'(\Phi)=0$, then there appears a stable local minimum at $\Phi_{0}$ such that $V(\Phi_{0}) \equiv \Lambda$ will be the value of an effective cosmological constant. Then, there exists a typical background Ricci scalar $R_{0} \sim \Lambda$. Around $R \sim R_{0}$, the $f(R)$ dark energy model (before cured) is a good description. On the other hand, due to the perturbation of the field during the gravitational collapses, $R$ can increase and if it reaches a certain bound, then $R + (c/2) R^{2}$ becomes a better approximation. At once the theory is approximated by the $R + (c/2) R^{2}$ model, which effectively restricts the increase of the Ricci scalar. Therefore approximately, the maximum possible Ricci scalar $R_{\mathrm{max}}$ occurs when the correction term $(c/2)R^{2}$ becomes comparable to the leading term $R$, or $\Lambda \simeq c R_{\mathrm{max}}^{2}$. In FIG.~\ref{fig:causal}, $\Lambda \sim R_{0} \sim 10^{-3}$ and $c \sim 10^{-3}$. Therefore, $R_{\mathrm{max}} \sim \mathcal{O}(1)$ and this is shown by FIG.~\ref{fig:Ricci}.

\begin{figure}
\begin{center}
\includegraphics[scale=0.5]{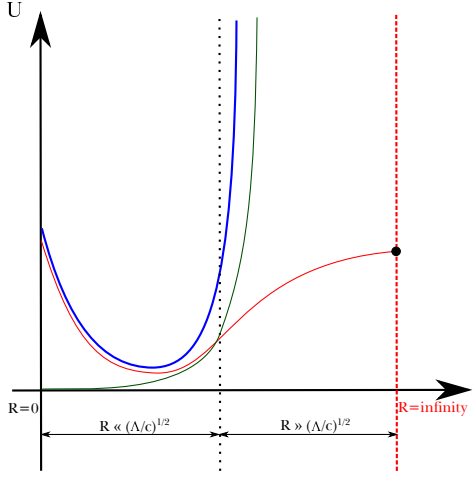}
\caption{\label{fig:concept}Conceptual diagram of the effective potential $U(R)$. The red curve corresponds to the $f(R)$ dark energy models with the cusp singularity; one can reach $R = \infty$ with a finite energy. The green curve corresponds to the $R + (c/2)R^{2}$ model. The blue curve is the cured model via the $R^{2}$ term, where it is similar as the red curve for the small curvature region and the green curve for the large curvature region. Between two regions, there is an effective boundary which is on the order of $R \sim (\Lambda/c)^{1/2}$.}
\end{center}
\end{figure}

FIG.~\ref{fig:concept} gives a conceptual picture. The $R^{2}$ term can cure the cusp singularity and hence we obtain the blue colored curve, where it follows the red colored curve around the small curvature regime and follows the green colored curve for the large curvature regime. Basically, $R \sim (\Lambda/c)^{1/2}$ becomes the boundary between two different regimes. The gravitational collapse perturb the scalar field away from the local minimum $R \ll (\Lambda/c)^{1/2}$ to become $R \sim (\Lambda/c)^{1/2}$.

Now the question is, what are the proper values of $c$? Note that the effective mass of the scalar field is $m^{2} = 1/c$. Therefore, there is a maximum bound of $c$ such that $c < m_{\mathrm{bd}}^{-2}$, where $m_{\mathrm{bd}}$ is the experimental bound of the scalar field mass. Note that this condition can be weakened by the Chameleon mechanism \cite{Khoury:2003aq}. As we also want to include the inflation scenario in the $f(R)$ gravity \cite{Starobinsky:1980te}, we find that the observational constraints restrict $m \simeq 10^{-5}$ in Planck units \cite{Ade:2015lrj}. Although this is not very conclusive, a reasonable interpretation is that, since $\Lambda \sim 10^{-123} \ll 1$ in terms of the Planck units, $R_{\mathrm{max}} \sim 10^{-66}$ would be way sub-Planckian (see also discussions in \cite{Dutta:2015nga}).

\subsection{Pros: horizon as a super-Planckian surface?}

If we regard $f(R)$ gravity as a possible modification of the Einstein gravity, then the cusp singularity appears inevitable from Ricci scalar terms. Of course, such behavior is generic and can be found not only for the Ricci scalar, but also other types of curvature terms, though these will be technically more difficult to investigate. If it is the case, then this means that a gravitational collapse can \textit{trigger} higher curvature corrections and this can cause a curvature singularity that is not in Einstein gravity. Such a curvature singularity can be located outside as well as inside the horizon ((A) and (B) of FIG.~\ref{fig:diagram}). The point is that the areal radius of the singularity is non-zero if such a singularity is induced by higher curvature terms.

Let us focus on the recent discussion on the information loss problem \cite{Hawking:1976ra}. According to the black hole complementarity argument \cite{Susskind:1993if}, the unitarity can be preserved for all processes of a black hole, from collapse to evaporation, to an asymptotic observer, as well as an in-falling observer based on quantum field theory of curved spacetime. However, later it was noticed that black hole complementarity is inconsistent \cite{Yeom:2008qw,Almheiri:2012rt} and hence we need to drop one of the assumptions. The authors in \cite{Almheiri:2012rt} tried to drop general relativity for an in-falling observer around the horizon scale, and they said that there exists a firewall around the horizon scale. If there exists a firewall and if it can prevent the problem of black hole complementarity, then it should affect to outside the event horizon \cite{Hwang:2012nn}.

If there exists a firewall, then it may resolve the tension of black hole complementarity; however, what is the mechanism to realize the firewall? Up to now, there is no consensus (for useful comments, see \cite{Sasaki:2014spa,Chen:2014jwq}). However, if we remind the $f(R)$ cusp singularities, then it opens a new interesting possibility. If there exists higher order curvature corrections from quantum gravitational effects, such effects can be triggered by gravitational collapses. Note that if it does not have a cusp singularity, then it will not be triggered by gravitational collapses, since gravitational collapses are in general sub-Planckian.

If there is a horizon scale curvature singularity that prevents an in-falling observer, can we identify this as a firewall? Of course, we do not have enough evidences whether a cusp singularity should appear for every cases. However, at least, this is one of logical possibilities; also, the properties (causal structures) of the cusp singularity induced by $f(R)$ seem to be quite consistent with the notion of the firewall. Therefore, it needs more investigations for future studies.

\section{\label{sec:dis}Discussion}

In this paper, we investigated gravitational collapses in $f(R)$ gravity. For some dark energy models, e.g., Starobinsky model \cite{Starobinsky:2007hu} and Hu-Sawicki model \cite{Hu:2007nk}, gravitational collapses can trigger to create a cusp singularity. Such a cusp singularity can be cured by adding a quadratic term $\sim c R^{2}$, but still there can exist a Ricci scalar bump around the horizon scale.

If one of $f(R)$ dark energy models is correct, then it is very natural to accept the existence of a cusp singularity, or at least, its cured effect as a Ricci scalar bump. Then we can ask this question: can we really observe such an effect outside the horizon?

Regarding this, one may discuss \textit{against} this: the curvature scale of a Ricci scalar bump will be approximately $R_{\mathrm{max}} \sim \sqrt{\Lambda/c}$, where $\Lambda$ is the cosmological constant of the current universe. Since our cosmological constant is much smaller than the Planck scale, i.e., $\Lambda \ll 1$, hence $R_{\mathrm{max}}$ will be much smaller than the Planck scale.

On the other hand, one may also discuss \textit{for} this: such a cusp singularity or a Ricci scalar bump can role as a modification of general relativity even around the horizon scale. Here, a cusp singularity is important; usually a gravitational collapsing process is sub-Planckian and hence its back-reaction to higher curvature corrections will be much weaker. However, if there exists a cusp singularity, then super-Planckian effect can be triggered by sub-Planckian gravitational collapses.

Can we identify this as a firewall \cite{Almheiri:2012rt} that resolves troubles of black hole complementarity \cite{Yeom:2008qw}? It is still unclear. At once there is a cusp singularity, even though it is inside the horizon, it will be useful to prevent inconsistency of black hole complementarity. In addition, the existence of a cusp singularity can be argued that it is originated from quantum corrections of gravity. However, the existence of a cusp singularity requires an exotic shape of terms, e.g., $R^{-n}$, while their theoretical stability is not clear.

At least, we open this as a logical possibility. Perhaps, a gravitational collapse can be a good window to investigate Planck scale physics. For further justifications or criticisms, we remain as a future work.

\newpage

\section*{Acknowledgment}
PC and DY are supported by National Center for Theoretical Sciences (NCTS) of Taiwan, Ministry of
Science and Technology (MOST) of Taiwan, and the Leung Center for Cosmology and Particle Astrophysics (LeCosPA) of
National Taiwan University.

\section*{\label{sec:appa}Appendix A. Implementation of double-null formalism}

The Einstein tensors and energy-momentum tensor components are expressed as follows ($\omega = 0$):
\begin{eqnarray}
\label{eq:Guu}G_{uu} &=& -\frac{2}{r} \left(f_{,u}-2fh \right),\\
\label{eq:Guv}G_{uv} &=& \frac{1}{2r^{2}} \left( 4 rf_{,v} + \alpha^{2} + 4fg \right),\\
\label{eq:Gvv}G_{vv} &=& -\frac{2}{r} \left(g_{,v}-2gd \right),\\
\label{eq:Gthth}G_{\theta\theta} &=& -4\frac{r^{2}}{\alpha^{2}} \left(d_{,u}+\frac{f_{,v}}{r}\right),\\
\label{eq:TBDuu}T^{\mathrm{BD}}_{uu} &=& \frac{1}{8 \pi \Phi} (W_{,u} - 2hW), \\
\label{eq:TBDuv}T^{\mathrm{BD}}_{uv} &=& - \frac{Z_{,u}}{8 \pi \Phi} - \frac{gW+fZ}{4 \pi r \Phi} + \frac{\alpha^{2}V}{32 \pi \Phi},\\
\label{eq:TBDvv}T^{\mathrm{BD}}_{vv} &=& \frac{1}{8 \pi \Phi} (Z_{,v} - 2dZ), \\
\label{eq:TBDthth}T^{\mathrm{BD}}_{\theta\theta} &=& \frac{r^{2}}{2 \pi \alpha^{2} \Phi} Z_{,u} + \frac{r}{4 \pi \alpha^{2} \Phi} (gW+fZ) - \frac{r^{2}V}{16 \pi \Phi},\\
\label{eq:TSuu}T^{\mathrm{C}}_{uu} &=& \frac{1}{4\pi} \left[ w\overline{w} + iea(\overline{w}s-w\overline{s}) +e^{2}a^{2}s\overline{s} \right],\\
\label{eq:TSuv}T^{\mathrm{C}}_{uv} &=& \frac{{(a_{,v})}^{2}}{4\pi\alpha^{2}},\\
\label{eq:TSvv}T^{\mathrm{C}}_{vv} &=& \frac{1}{4\pi} z\overline{z},\\
\label{eq:TSthth}T^{\mathrm{C}}_{\theta\theta} &=& \frac{r^{2}}{4\pi\alpha^{2}} \left[ (w\overline{z}+z\overline{w}) + iea(\overline{z}s-z\overline{s})+\frac{2{(a_{,v})}^{2}}{\alpha^{2}} \right].
\end{eqnarray}

After solving a coupled system of equations, we can write all the equations as follows:
\begin{eqnarray}
\label{eq:E1}f_{,u} &=& 2fh - \frac{r}{2 \Phi} (W_{,u}-2hW) - \frac{4 \pi r}{\Phi} T^{\mathrm{C}}_{uu} ,\\
\label{eq:E2}g_{,v} &=& 2gd - \frac{r}{2 \Phi} (Z_{,v}-2dZ) - \frac{4 \pi r}{\Phi} T^{\mathrm{C}}_{vv} ,\\
\label{eq:E3}d_{,u} = h_{,v} &=& \mathfrak{A} - \frac{\mathfrak{B}}{r} - \frac{\mathfrak{C}}{2r\Phi},\\
\label{eq:E4}g_{,u} = f_{,v} &=& \mathfrak{B} - \frac{\mathfrak{C}}{2\Phi},\\
\label{eq:Phi}Z_{,u} = W_{,v} &=& \frac{\mathfrak{C}}{r},
\end{eqnarray}
where
\begin{eqnarray}
\label{eq:A}\mathfrak{A} &\equiv&-\frac{2\pi \alpha^{2}}{r^{2}\Phi}T^{\mathrm{C}}_{\theta\theta} - \frac{1}{2r}\frac{1}{\Phi}(gW+fZ) + \frac{\alpha^{2}V}{8 \Phi}, \\
\label{eq:B}\mathfrak{B} &\equiv& \frac{4 \pi r}{\Phi}T^{\mathrm{C}}_{uv} - \frac{\alpha^{2}}{4r} - \frac{fg}{r} - \frac{1}{\Phi}(gW+fZ) + \frac{r \alpha^{2}}{8 \Phi}V, \\
\label{eq:C}\mathfrak{C} &\equiv& - fZ - gW - \frac{4\pi r}{3} \left(\frac{\alpha^{2}}{2}T^{\mathrm{C}} + \frac{\alpha^{2}}{16 \pi}\left(\Phi V' - 2 V \right) \right),
\end{eqnarray}
as well as matter field equations
\begin{eqnarray}
\label{eq:fieldeqns1}a_{,v} &=& \frac{\alpha ^{2} q}{2 r^{2}}, \\
\label{eq:fieldeqns2}q_{,v} &=& -\frac{ier^{2}}{2} (\overline{s}z-s\overline{z}), \\
\label{eq:fieldeqns3}z_{,u} = w_{,v} &=& - \frac{fz}{r} - \frac{gw}{r} - \frac{iearz}{r} - \frac{ieags}{r} - \frac{ie}{4r^{2}}\alpha^{2}qs.
\end{eqnarray}
In addition, in order to avoid a difficulty to convert $f(R)$ to $V(\Phi)$, we substitute using the following identities:
\begin{eqnarray}
V(\Phi) &=& -f(R) + R f'(R),\\
\Phi \frac{dV}{d\Phi} - 2 V &=& 2f(R) - R f'(R).
\end{eqnarray}
Therefore, we need an evolution of the Ricci scalar $R$. To calculate the evolution of $R$, we further use the following identities:
\begin{eqnarray}\label{eq:R}
R_{,u} = \frac{\Phi_{,u}}{f''(R)}, \;\;\; R_{,v} = \frac{\Phi_{,v}}{f''(R)}.
\end{eqnarray}

Now, equations for $\alpha_{,uv}$, $r_{,uv}$, $\Phi_{,uv}$, and $s_{,uv}$ can be represented by a set of first order differential equations.
We use the same integration method that was used in previous papers \cite{Hwang:2011kg,Hong:2008mw,Hansen:2009kn,Hwang:2010aj}.
We use the second order Runge-Kutta method \cite{nr}.

\section*{\label{sec:appb}Appendix B. Consistency and convergence tests}

In this appendix, we report on the convergence and consistency tests for our simulations. We tested for $f_{\mathrm{cure,S}}(R)$ with ($A=0.25$, $e=0.3$, $c=0.001$) and $f_{\mathrm{cure,HS}}(R)$ with ($A=0.25$, $e=0.6$ $c=0.001$), since these cases include all the interesting features of the present paper.

\begin{figure}
\begin{center}
\includegraphics[scale=0.25]{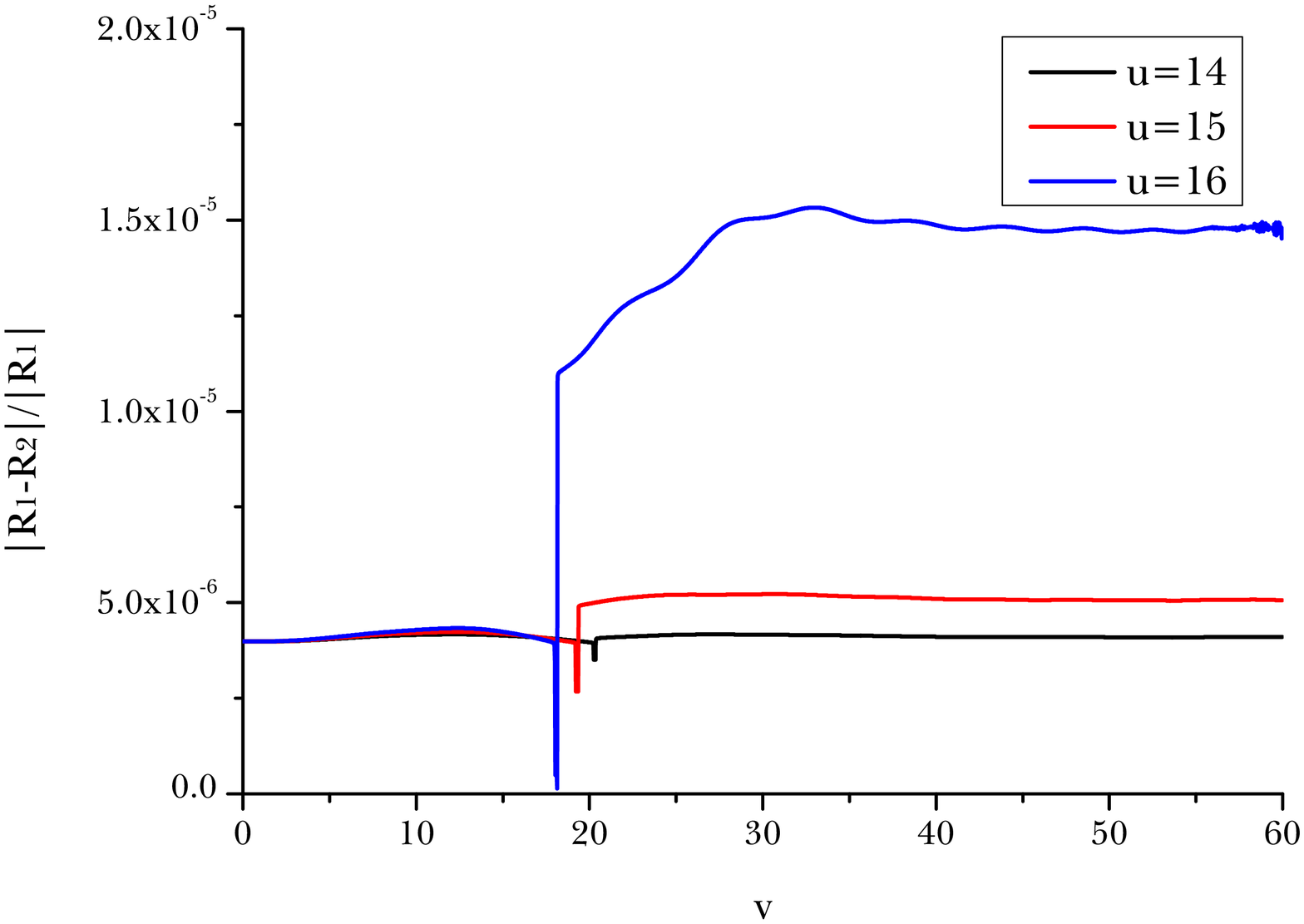}
\includegraphics[scale=0.25]{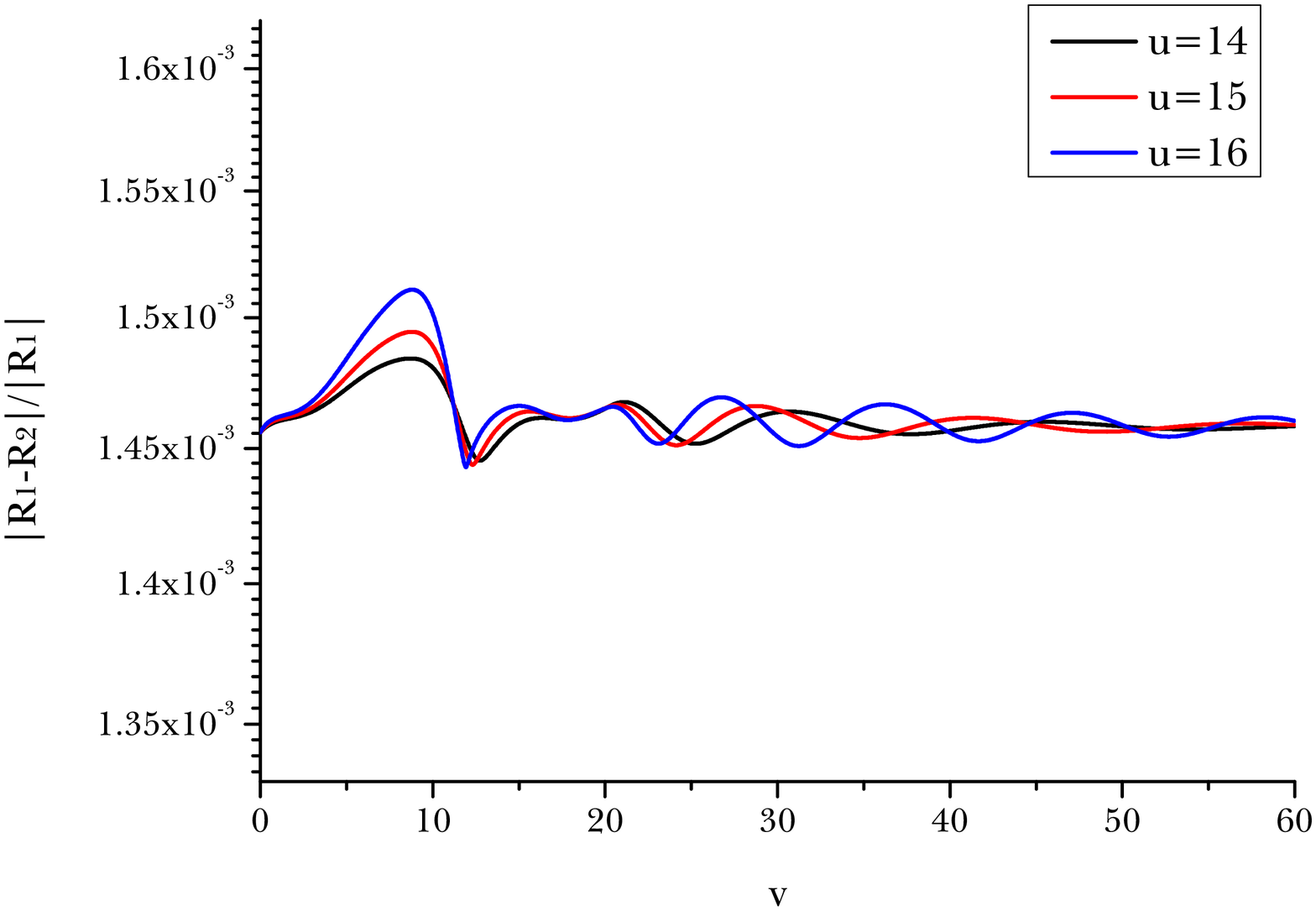}
\caption{\label{fig:consistency}Consistency test for $|R_{1}-R_{2}|/|R_{1}|$ around $u=14, 15, 16$ for $f_{\mathrm{cure,S}}(R)$ (Left) and $f_{\mathrm{cure,HS}}(R)$ (Right).}
\end{center}
\end{figure}

For consistency, we can check various relations, but the most important non-trivial test for $f(R)$ gravity will be the Ricci scalar. We calculated it by using Eq.~(\ref{eq:R}), while the definition of the Ricci scalar is
\begin{eqnarray}\label{eq:Ricci}
R = \frac{2}{\alpha^{2}} \left[ 4 \left(\frac{\alpha_{,u}}{\alpha}\right)_{,v} + 8 \frac{r_{,uv}}{r} + \frac{\alpha^{2}}{r^{2}} + \frac{4r_{,u}r_{,v}}{r^{2}} \right].
\end{eqnarray}
We call the latter $R_{1}$ and the former $R_{2}$, and checked $|R_{1}-R_{2}|/|R_{1}|$ around $u=14, 15, 16$. FIG.~\ref{fig:consistency} shows that the differences are less than $10^{-3}$~\% for the Starobinsky model and $10^{-1}$~\% for the Hu-Sawicki model.

\begin{figure}
\begin{center}
\includegraphics[scale=0.25]{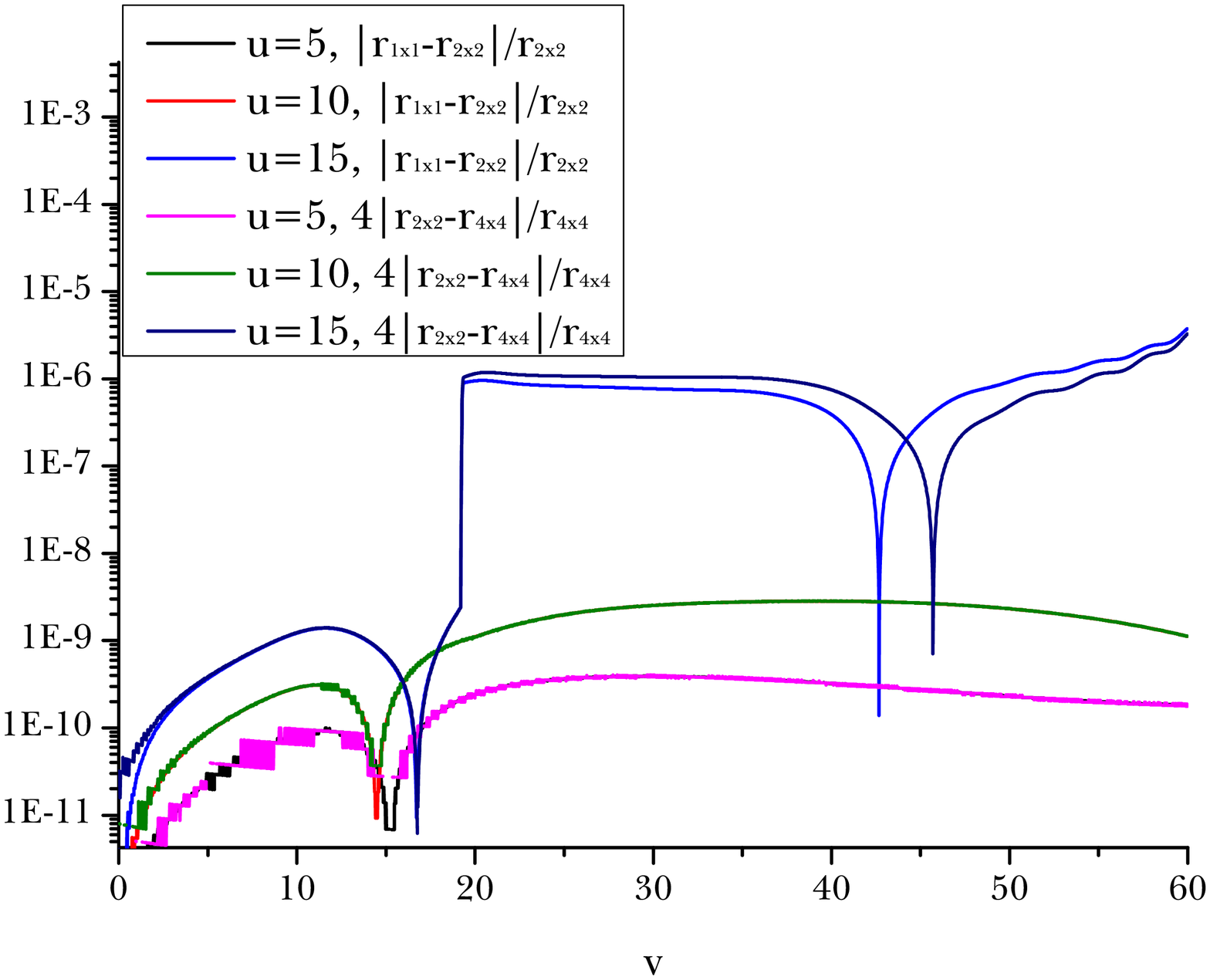}
\includegraphics[scale=0.25]{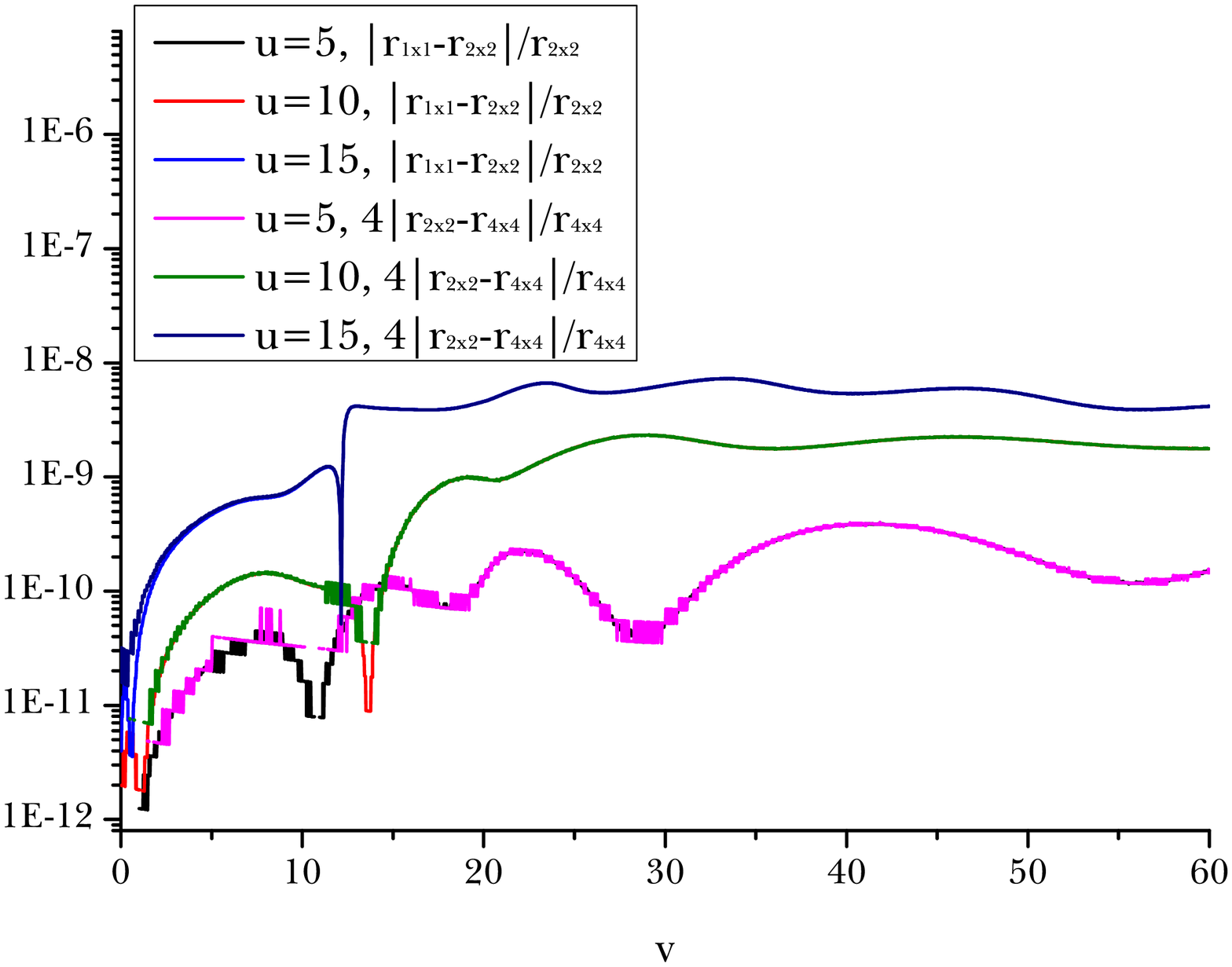}
\caption{\label{fig:convergence}Convergence test: $|r_{1\times1}-r_{2\times2}|/r_{2\times2}$ and $4|r_{2\times2}-r_{4\times4}|/r_{4\times4}$ around $u=5, 10, 15$ for $f_{\mathrm{cure,S}}(R)$ (Left) and $f_{\mathrm{cure,HS}}(R)$ (Right). This shows the second order convergence.}
\end{center}
\end{figure}

For convergence, we compared finer simulations: $1\times1$, $2\times2$, and $4\times4$ times finer. In FIG.~\ref{fig:convergence}, we see that the difference between the $1\times1$ and $2\times2$ times finer cases is $4$ times the difference between the $2\times2$ and $4\times4$ times finer cases, and thus our simulation converges to second order. The numerical error is less than $10^{-4}$~\% for the Starobinsky model and $10^{-6}$~\% for the Hu-Sawicki model.



\newpage

\begin{thebibliography}{200}

\bibitem{BBS}
  K. Becker, M. Becker and J. H. Schwarz, {\it ``String theory and M-theory: A modern introduction,''} Cambridge University Press (2007).

\bibitem{Gasperini:2007zz}
  M.~Gasperini, {\it ``Elements of string cosmology,''} Cambridge University Press (2007).

\bibitem{ST}
  Y. Fujii and K. Maeda, {\it ``The scalar-tensor theory of gravitation,''} Cambridge University Press (2003); \\
  V. Faraoni, {\it ``Cosmology in scalar tensor gravity,''} Kluwer Academic Publishers (2004).

\bibitem{Sotiriou:2008rp}
  T.~P.~Sotiriou and V.~Faraoni,
  Rev.\ Mod.\ Phys.\  {\bf 82}, 451 (2010)
  [arXiv:0805.1726 [gr-qc]];\\
  S.~Nojiri and S.~D.~Odintsov,
  eConf {\bf C0602061}, 06 (2006)
  [Int.\ J.\ Geom.\ Meth.\ Mod.\ Phys.\  {\bf 4}, 115 (2007)]
  [arXiv:hep-th/0601213].

\bibitem{Brans:1961sx}
  C.~Brans and R.~H.~Dicke,
  Phys.\ Rev.\  {\bf 124}, 925 (1961).

\bibitem{Starobinsky:1980te}
  A.~A.~Starobinsky,
  Phys.\ Lett.\ B {\bf 91}, 99 (1980).

\bibitem{Ade:2015lrj} 
  P.~A.~R.~Ade {\it et al.}  [Planck Collaboration],
  arXiv:1502.02114 [astro-ph.CO].

\bibitem{Starobinsky:2007hu} 
  A.~A.~Starobinsky,
  JETP Lett.\  {\bf 86}, 157 (2007)
  [arXiv:0706.2041 [astro-ph]].

\bibitem{Hu:2007nk} 
  W.~Hu and I.~Sawicki,
  Phys.\ Rev.\ D {\bf 76}, 064004 (2007)
  [arXiv:0705.1158 [astro-ph]].

\bibitem{Appleby:2009uf} 
  S.~A.~Appleby, R.~A.~Battye and A.~A.~Starobinsky,
  JCAP {\bf 1006}, 005 (2010)
  [arXiv:0909.1737 [astro-ph.CO]].

\bibitem{Hwang:2011kg} 
  D.~Hwang, B.~H.~Lee and D.~Yeom,
  JCAP {\bf 1112}, 006 (2011)
  [arXiv:1110.0928 [gr-qc]].

\bibitem{Guo:2013dha} 
  J.~Q.~Guo, D.~Wang and A.~V.~Frolov,
  Phys.\ Rev.\ D {\bf 90}, no. 2, 024017 (2014)
  [arXiv:1312.4625 [gr-qc]].

\bibitem{Hamade:1995ce} 
  R.~S.~Hamade and J.~M.~Stewart,
  Class.\ Quant.\ Grav.\  {\bf 13}, 497 (1996)
  [gr-qc/9506044].

\bibitem{Hong:2008mw} 
  S.~E.~Hong, D.~Hwang, E.~D.~Stewart and D.~Yeom,
  Class.\ Quant.\ Grav.\  {\bf 27}, 045014 (2010)
  [arXiv:0808.1709 [gr-qc]];\\
  D.~Hwang and D.~Yeom,
  Phys.\ Rev.\ D {\bf 84}, 064020 (2011)
  [arXiv:1010.2585 [gr-qc]];\\
  D.~Hwang, H.~B.~Kim and D.~Yeom,
  Class.\ Quant.\ Grav.\  {\bf 29}, 055003 (2012)
  [arXiv:1105.1371 [gr-qc]];\\
  J.~Hansen, B.~H.~Lee, C.~Park and D.~Yeom,
  Class.\ Quant.\ Grav.\  {\bf 30}, 235022 (2013)
  [arXiv:1307.0266 [hep-th]].

\bibitem{doublenull}
  R.~Parentani and T.~Piran, Phys.\ Rev.\ Lett {\bf 73}, 2805 (1994) [arXiv:hep-th/9405007];\\
  S.~Ayal and T.~Piran, Phys.\ Rev.\  D {\bf 56}, 4768 (1997) [arXiv:gr-qc/9704027];\\
  S.~Hod and T.~Piran, Phys.\ Rev.\ Lett {\bf 81}, 1554 (1998) [arXiv:gr-qc/9803004];\\
  S.~Hod and T.~Piran, Gen.\ Rel.\ Grav.\  {\bf 30}, 1555 (1998) [arXiv:gr-qc/9902008];\\
  E.~Sorkin and T.~Piran, Phys.\ Rev.\ D {\bf 63}, 084006 (2001) [arXiv:gr-qc/0009095];\\
  E.~Sorkin and T.~Piran, Phys.\ Rev.\  D {\bf 63}, 124024 (2001) [arXiv:gr-qc/0103090];\\
  Y.~Oren and T.~Piran, Phys.\ Rev.\ D {\bf 68}, 044013 (2003) [arXiv:gr-qc/0306078];\\
  J.~Hansen, A.~Khokhlov and I.~Novikov, Phys.\ Rev.\  D {\bf 71}, 064013 (2005) [arXiv:gr-qc/0501015];\\
  A.~Doroshkevich, J.~Hansen, I.~Novikov and A.~Shatskiy, Int.\ J.\ Mod.\ Phys.\ D {\bf 18}, 1665 (2009) [arXiv:0812.0702 [gr-qc]]; \\
  P.~P.~Avelino, A.~J.~S.~Hamilton and C.~A.~R.~Herdeiro, Phys.\ Rev.\  D {\bf 79}, 124045 (2009) [arXiv:0904.2669 [gr-qc]]; \\
  A.~Doroshkevich, J.~Hansen, D.~Novikov, I.~Novikov and A.~Shatskiy, Phys.\ Rev.\  D {\bf 81}, 124011 (2010)
  [arXiv:0908.1300 [gr-qc]].

\bibitem{Hansen:2009kn} 
  J.~Hansen, D.~Hwang and D.~Yeom,
  JHEP {\bf 0911}, 016 (2009)
  [arXiv:0908.0283 [gr-qc]];\\
  D.~Hwang and D.~Yeom,
  Class.\ Quant.\ Grav.\  {\bf 28}, 155003 (2011)
  [arXiv:1010.3834 [gr-qc]];\\
  D.~Hwang, B.~H.~Lee, W.~Lee and D.~Yeom,
  JCAP {\bf 1207}, 003 (2012)
  [arXiv:1201.6109 [gr-qc]].

\bibitem{Scheel:1994yr} 
  M.~A.~Scheel, S.~L.~Shapiro and S.~A.~Teukolsky,
  Phys.\ Rev.\ D {\bf 51}, 4208 (1995)
  [gr-qc/9411025];\\
  M.~A.~Scheel, S.~L.~Shapiro and S.~A.~Teukolsky,
  Phys.\ Rev.\ D {\bf 51}, 4236 (1995)
  [gr-qc/9411026];\\
  T.~Chiba and J.~Soda,
  Prog.\ Theor.\ Phys.\  {\bf 96}, 567 (1996)
  [gr-qc/9603056];\\
  A.~Borkowska, M.~Rogatko and R.~Moderski,
  Phys.\ Rev.\ D {\bf 83}, 084007 (2011)
  [arXiv:1103.4808 [hep-th]];\\
  A.~Nakonieczna, M.~Rogatko and R.~Moderski,
  Phys.\ Rev.\ D {\bf 86}, 044043 (2012)
  [arXiv:1209.1203 [hep-th]];\\
  A.~Nakonieczna and M.~Rogatko,
  Gen.\ Rel.\ Grav.\  {\bf 44}, 3175 (2012)
  [arXiv:1209.3614 [hep-th]].


\bibitem{Hwang:2010aj} 
  D.~Hwang and D.~Yeom,
  Class.\ Quant.\ Grav.\  {\bf 27}, 205002 (2010)
  [arXiv:1002.4246 [gr-qc]];\\
  D.~Hwang, F.~G.~Pedro and D.~Yeom,
  JHEP {\bf 1309}, 159 (2013)
  [arXiv:1306.6687 [hep-th]];\\
  J.~Hansen and D.~Yeom,
  JHEP {\bf 1410}, 40 (2014)
  [arXiv:1406.0976 [hep-th]];\\
  J.~Hansen and D.~Yeom,
  arXiv:1506.05689 [hep-th].

\bibitem{Khoury:2003aq}
  J.~Khoury and A.~Weltman,
  Phys.\ Rev.\ Lett.\  {\bf 93}, 171104 (2004)
  [arXiv:astro-ph/0309300].

\bibitem{Dutta:2015nga} 
  K.~Dutta, S.~Panda and A.~Patel,
  arXiv:1504.05790 [gr-qc].

\bibitem{Hawking:1976ra}
  S.~W.~Hawking,
  Phys.\ Rev.\  D {\bf 14}, 2460 (1976).

\bibitem{Susskind:1993if}
  L.~Susskind, L.~Thorlacius and J.~Uglum,
  Phys.\ Rev.\  D {\bf 48}, 3743 (1993)
  [arXiv:hep-th/9306069].

\bibitem{Yeom:2008qw}
  D.~Yeom and H.~Zoe,
  Phys.\ Rev.\  D {\bf 78}, 104008 (2008)
  [arXiv:0802.1625 [gr-qc]]; \\
  D.~Yeom,
  Int.\ J.\ Mod.\ Phys.\ CS {\bf 1}, 311 (2011) [arXiv:0901.1929 [gr-qc]];\\
  D.~Yeom and H.~Zoe,
  Int.\ J.\ Mod.\ Phys.\ A {\bf 26}, 3287 (2011)
  [arXiv:0907.0677 [hep-th]];\\
  P.~Chen, Y.~C.~Ong and D.~Yeom,
  JHEP {\bf 1412}, 021 (2014)
  [arXiv:1408.3763 [hep-th]].

\bibitem{Almheiri:2012rt} 
  A.~Almheiri, D.~Marolf, J.~Polchinski and J.~Sully,
  JHEP {\bf 1302}, 062 (2013)
  [arXiv:1207.3123 [hep-th]];\\
  A.~Almheiri, D.~Marolf, J.~Polchinski, D.~Stanford and J.~Sully,
  JHEP {\bf 1309}, 018 (2013)
  [arXiv:1304.6483 [hep-th]].

\bibitem{Hwang:2012nn} 
  D.~Hwang, B.~-H.~Lee and D.~Yeom,
  JCAP {\bf 1301}, 005 (2013)
  [arXiv:1210.6733 [gr-qc]];\\
  W.~Kim, B.~-H.~Lee and D.~Yeom,
  JHEP {\bf 1305}, 060 (2013)
  [arXiv:1301.5138 [gr-qc]];\\
  B.~-H.~Lee and D.~Yeom,
  Nucl.\ Phys.\ Proc.\ Suppl.\  {\bf 246-247}, 178 (2014)
  [arXiv:1302.6006 [gr-qc]].

\bibitem{Sasaki:2014spa} 
  M.~Sasaki and D.~Yeom,
  JHEP {\bf 1412}, 155 (2014)
  [arXiv:1404.1565 [hep-th]];\\
  B.~H.~Lee, W.~Lee and D.~Yeom,
  arXiv:1502.07471 [hep-th].

\bibitem{Chen:2014jwq} 
  P.~Chen, Y.~C.~Ong and D.~Yeom,
  arXiv:1412.8366 [gr-qc].

\bibitem{nr}
  W.~H.~Press, S.~A.~Teukolsky, W.~T.~Vetterling and B.~P.~Flannery, {\it ``Numerical Recipes: The Art of Scientific Computing,'' 3rd ed.,}
  Cambridge University Press (2007).




\end{thebibliography}
\end{document}